# Title: Universal Synthesis of Large-Area Hybrid Perovskite Monolayers Enabling Lithography-Free, Low-Static-Power Optoelectronics

**Authors:** Shankeerthan Kalyanasundaram[1], Swarit Dwivedi[1], Sanje Mahasivam[2], Christopher R. McNeill[1], Sudha Mokkapati[1*]

**Affiliations:**

[1]Department of Materials Science and Engineering, Monash University, Clayton, 3800, Australia

[2]School of Science, RMIT University, Melbourne, 3000, Australia

*Corresponding author. Email: Sudha.mokkapati@monash.edu

**Abstract:**

Solution-phase synthesis of monolayer hybrid organic-inorganic perovskites is constrained by rapid out-of-plane crystal growth, uncontrolled nucleation, and strong interlayer aggregation, yielding flakes of only a few µm. We introduce a steric-engineering ternary-solvent growth strategy to synthesize monolayer and few-layer $A_2MX_4$ hybrid perovskites with exceptionally large surface areas. By combining alcohol cosolvents with an amine additive, out-of-plane stacking is suppressed. Using $(PEA)_2CuCl_4$ as a model system, monolayers with lateral areas exceeding 1000 $\mu m^2$ are achieved, representing a massive dimensional increase over conventional methods. Molecular dynamics and quantum-chemical calculations show that longer-chain alcohols reorganize precursor solvation, while solvent volatility prolongs edge-mediated precursor transport during annealing. This strategy demonstrates universal applicability across diverse organic spacer cations ($PEA^+$, F-$PEA^+$, $PentA^+$), metal centers ($Cu^{2+}$, $Mn^{2+}$, $Pb^{2+}$), and halides ($Cl^-$, $Br^-$, $I^-$). The large lateral dimensions enable direct, solvent-free electrical contact fabrication via physical shadow masking, bypassing chemical degradation from solvents used in conventional lithography. Planar devices fabricated from isolated ultrathin $(PEA)_2PbI_4$ channels exhibit ultra-low dark currents down to the $10^{-15}$–$10^{-13}$A range, enabling femtowatt-to-picowatt static power consumption, ~ 0.2 pW at 5 V, a 250-fold reduction relative to drop-cast films, while maintaining a high light-to-dark current ratio of $4.3 \times 10^2$. This work establishes precise steric and solvation control as a paradigm for scalable, compositionally diverse, low-static-power monolayer 2D perovskite optoelectronics.

**Main Text:**

Organic-inorganic hybrid perovskites have attracted widespread attention over the past decade due to their outstanding optoelectronic performance, particularly in photovoltaics[1,2]. Within this materials family, layered two-dimensional (2D) hybrid perovskites have emerged as an important class of low-dimensional semiconductors owing to their tunable optoelectronic properties, improved environmental stability, and structural versatility[3,4]. These materials consist of inorganic metal–halide layers separated by organic spacer cations and are commonly described by the formula $A_2MX_4$ in their pure 2D structure (**Fig. 1a**), where A is a monovalent organic spacer cation, M is a divalent metal cation, and X is a halide anion[5].

In this layered architecture, the metal-halide framework primarily governs the electronic and optical response, while the organic spacer modulates these properties through its influence on interlayer spacing, dielectric confinement, lattice distortion, and molecular packing.[6] While bulk layered perovskites inherently possess quantum-confined inorganic sheets[7–9], isolating these materials down to true single layers unlocks a fundamentally distinct physical regime. The transition to an isolated monolayer breaks out-of-plane structural symmetry and eliminates neighbouring dielectric screening, inducing giant dielectric contrast with the surrounding environment[10]. This drastic environmental change triggers uninhibited lattice relaxation, modified exciton-phonon coupling, and starkly altered exciton recombination dynamics that are strictly masked in bulk crystals[11,12]. Reaching high-quality, large-area monolayers is therefore not merely an incremental dimensional reduction, but an essential gateway to uncovering unscreened low-dimensional physics and tailoring next-generation single-layer optoelectronics.

Hybrid perovskite monolayers and few-layer flakes have mainly been obtained using mechanical exfoliation and liquid-phase synthesis[13–15]. Mechanical exfoliation can produce ultrathin flakes, but the resulting flakes are often irregular in shape, randomly distributed, variable in thickness and remains a low-throughput approach[12,13]. Liquid-phase synthesis offers a more direct route by forming monolayer and few-layer flakes directly on substrates under controlled crystallization conditions but existing methods commonly suffer from uncontrolled nucleation, aggregation, limited lateral size, and low yields of monolayer and few-layer crystals[14]. The ternary co-solvent method introduced by Dou et al. enabled synthesis of mono and few layer hybrid perovskite flakes, but the resulting ultrathin $(C_4H_9NH_3)_2PbBr_4$, square flakes were limited to micrometre-scale lateral dimensions, with edge lengths ranging between 1–10 µm and an average of ~ 4.2 µm[14]. Consequently, a universal solution-phase strategy that can reproducibly yield laterally large, isolated monolayer and few-layer crystals across diverse hybrid perovskite compositions remains lacking. Achieving this requires precise control over the crystallization environment to favour lateral crystal growth while suppressing excessive nucleation, out-of-plane growth, and aggregation. Previous studies of ternary-solvent-mediated hybrid perovskite crystal growth have shown that the constituent solvents can play distinct roles in precursor dissolution, supersaturation, nucleation, and anisotropic crystal growth, highlighting solvent selection as an important means of controlling crystal morphology.[16]

Here, we introduce a ternary-solvent-based steric-engineering strategy for growing laterally extended monolayer and few-layer hybrid perovskite crystals. The approach combines alcohol cosolvents of different chain lengths with a long-chain amine additive to regulate nucleation, aggregation, and out-of-plane growth. Using bis(phenethylammonium) tetrachlorocuprate, $(PEA)_2CuCl_4$, as a model system, we obtain well-isolated monolayers with lateral areas exceeding 1000 $\mu m^2$, substantially larger than those obtained using conventional solution-based approaches. We further demonstrate the generality of this strategy across $A_2MX_4$ hybrid perovskites containing different metal centres ($Cu^{2+}$, $Mn^{2+}$, and $Pb^{2+}$), organic spacer cations ($PEA^+$, $F\text{-}PEA^+$, and $PentA^+$), and halide anions ($Cl^-$, $Br^-$, and $I^-$). These results establish steric control of the crystallization environment as a versatile strategy for producing large-area, compositionally diverse hybrid perovskite monolayers and few-layer crystals.

Beyond synthesis, achieving large lateral dimensions is essential for translating these ultrathin crystals into functional optoelectronic devices. Standard device fabrication on small-area flakes relies heavily on lithography; however, hybrid perovskites are notoriously vulnerable to solvent degradation, structural damage, and chemical instability when exposed to conventional resists, developers, and strippers. The laterally extended nature of our solution-grown flakes circumvents this long-standing processing bottleneck by enabling direct, solvent-free electrical contact fabrication via physical shadow masking. To demonstrate the practical utility of this approach, we fabricate planar photoconductive channels using isolated ultrathin $(PEA)_2PbI_4$ flakes as active channels.

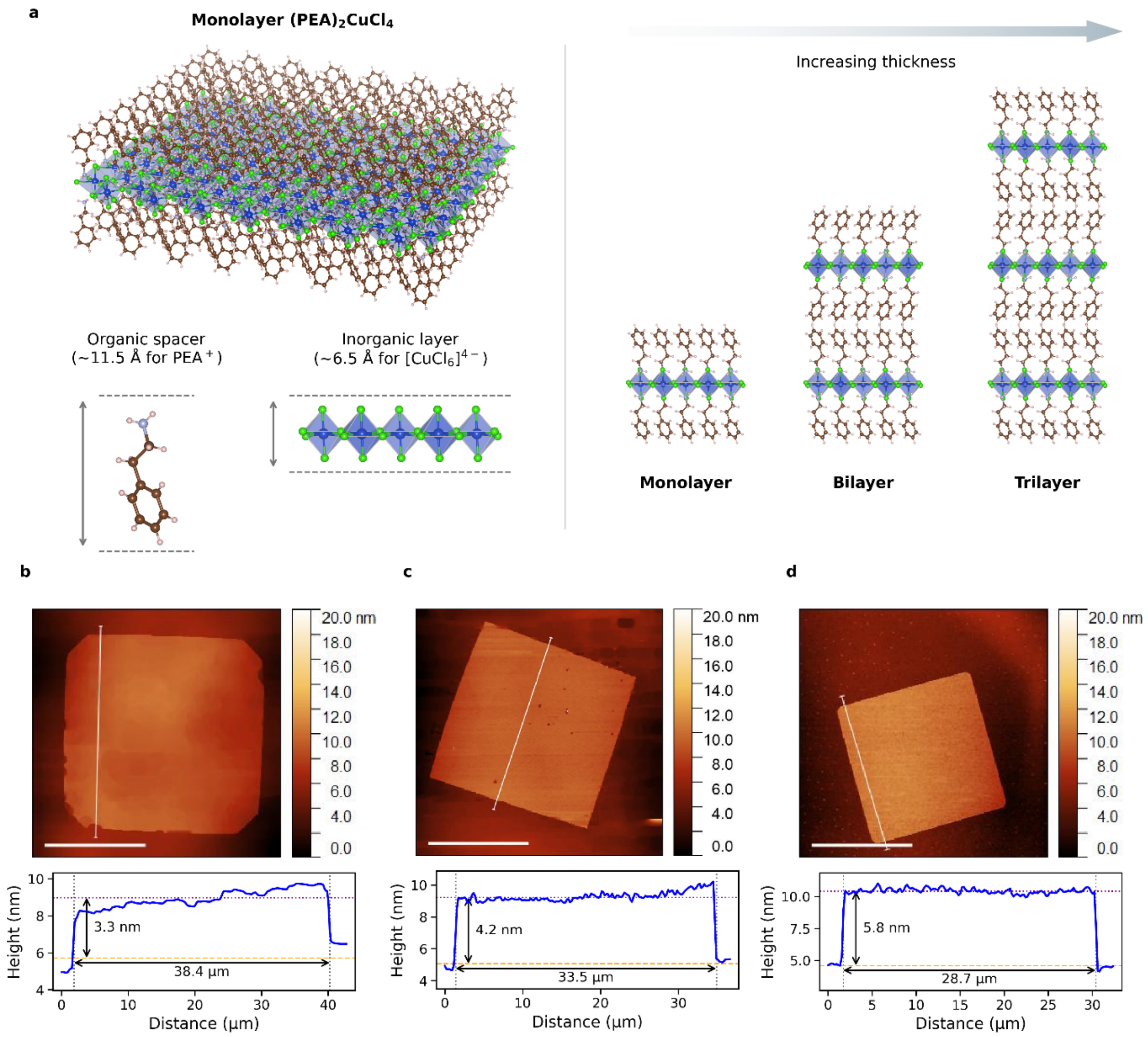

**Fig. 2: Large-area monolayer and few-layer 2D hybrid perovskites. (a)** Schematic illustration of monolayer, bilayer, and trilayer 2D hybrid perovskite structures composed of inorganic metal–halide layers separated by organic spacer cations. **(b-d)** Atomic force microscopy (AFM) topography images and corresponding height profiles of solution-grown monolayer, bilayer, and trilayer $(PEA)_2CuCl_4$ crystals, with thicknesses of approximately 3.3, 4.2, and 5.8 nm, respectively. Scale bars: 20 µm.

We synthesized 2D $(PEA)_2CuCl_4$ hybrid perovskite ultrathin crystals using a ternary-solvent approach. Phenethylammonium chloride (PEACl) and copper chloride ($CuCl_2$) were first dissolved in dimethylformamide (DMF) to prepare a precursor stock solution. A small aliquot of this solution was then introduced into a binary solvent mixture of chlorobenzene and alcohol to form the ternary growth solution. The mixture was then left undisturbed for 24 hours to allow the precursor-solvent system to equilibrate and develop prior to deposition. The resulting solution was drop-cast onto $SiO_2$ on Si substrate and annealed until complete solvent evaporation, yielding ultrathin $(PEA)_2CuCl_4$ flakes on the substrate surface. Full experimental details are provided in the **Methods section** of **Supplementary Information**.

**Fig. 2b-d** show AFM topography and the corresponding height profiles of monolayer, bilayer, and trilayer $(PEA)_2CuCl_4$ crystals. The AFM reveals step heights of 3.3, 4.2, and 5.8 nm, respectively. The measured thickness of the monolayer (3.3 nm) exceeds the bulk crystallographic interlayer spacing of ~ 1.9 nm[17]. This initial height offset is a well-known phenomenon for ultrathin 2D hybrid perovskites on $SiO_2/Si$ substrates, primarily caused by interfacial adsorbates, structural relaxation of the organic spacers in the single-layer limit, and substrate-dependent AFM tip-sample interactions[11,18].

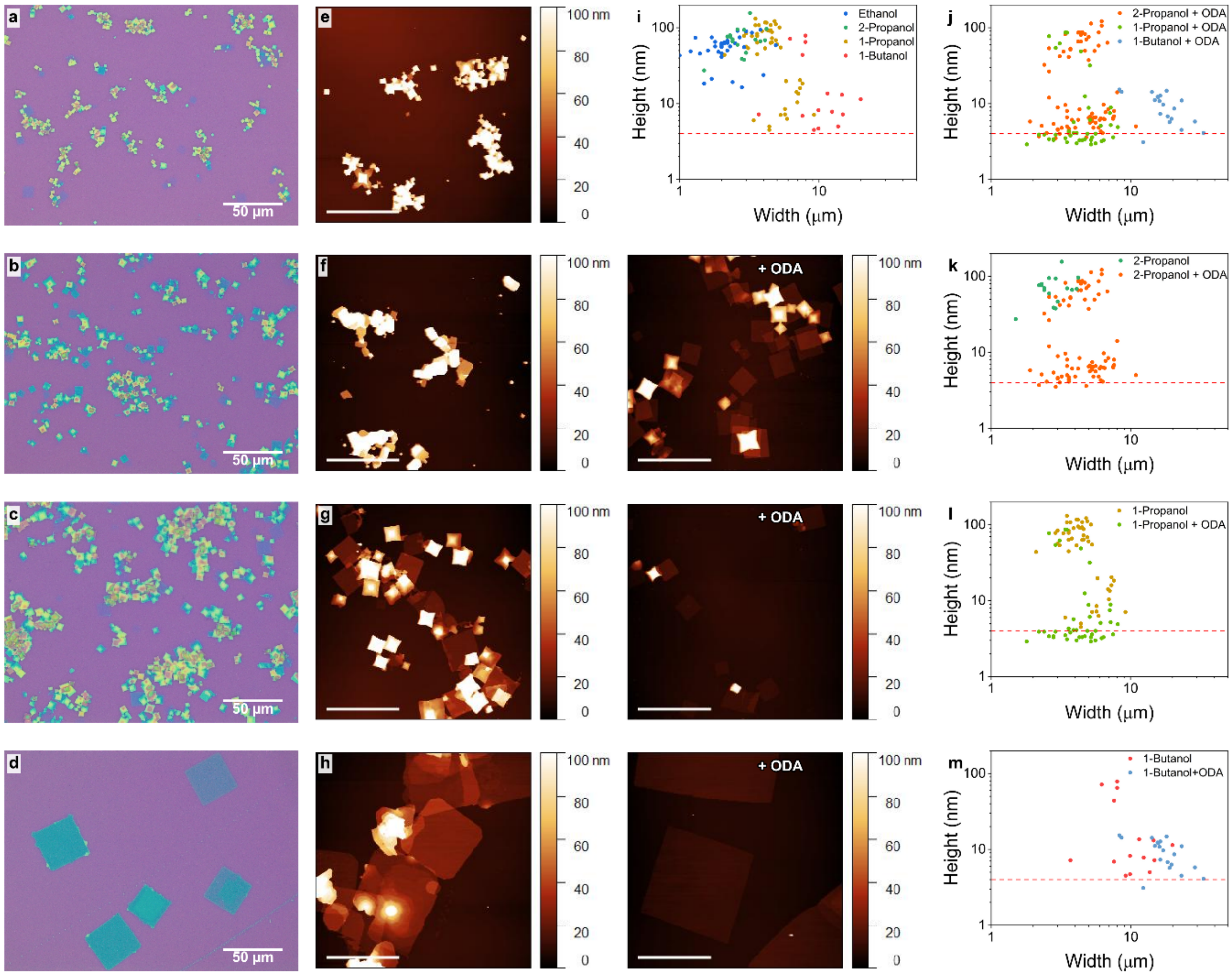

**Fig. 3: Effect of alcohol cosolvents and octadecylamine (ODA) as a steric additive on the morphology of $(PEA)_2CuCl_4$ flakes. (a–d)** Optical microscopy images of $(PEA)_2CuCl_4$ flakes prepared using a ternary solvent system (chlorobenzene:alcohol:DMF = 3200:400:10 by volume) with different alcohol cosolvents: **(a)** ethanol, **(b)** 2-propanol, **(c)** 1-propanol, and **(d)** 1-butanol. **(e)** AFM image of flakes synthesized using ethanol. **(f-h)** AFM images of flakes synthesized using **(f)** 2-propanol, **(g)** 1-propanol, and **(h)** 1-butanol, shown without ODA (left) and with ODA (right). **(i)** Combined height-width distributions of flakes

synthesized without ODA using ethanol, 2-propanol, 1-propanol, and 1-butanol, illustrating the effect of the alcohol cosolvent on flake morphology. **(j)** Combined height-width distributions of flakes synthesized with ODA using 2-propanol, 1-propanol, and 1-butanol. **(k-m)** Height-width distributions comparing flakes synthesized with and without ODA using **(k)** 2-propanol, **(l)** 1-propanol, and **(m)** 1-butanol, respectively, highlighting the influence of ODA on flake thickness and lateral dimensions for each solvent system. Red dashed lines in **(i-m)** indicate the estimated relaxed monolayer thickness based on the molecular structure shown in **Fig. 4a**. Scale bars in the AFM images: 20 µm.

**Fig. 5** demonstrates the influence of the alcohol cosolvent and octadecylamine (ODA) additive on the lateral and out-of-plane growth of $(PEA)_2CuCl_4$ flakes. Different alcohols were introduced into the ternary solvent system while maintaining the overall chlorobenzene:alcohol:DMF ratio at 3200:400:10 by volume. The alcohol carbon-chain length was systematically varied to investigate its influence on the lateral dimensions and thickness of the resulting perovskite flakes. Optical microscopy images of flakes synthesized without ODA are shown in **Fig. 6a–d**, while the corresponding AFM morphology is presented in **Fig. 7e–h**. For 2-propanol, 1-propanol, and 1-butanol, representative AFM images obtained in the presence of ODA are shown alongside the corresponding ODA-free samples in **Fig. 8f–h**, allowing direct comparison of the resulting morphologies. The height and lateral dimensions of individual flakes were extracted from AFM scans on individual flakes (**Supplementary Fig. 3**) to obtain representative height–width distributions. The combined distributions for samples synthesized without and with ODA are shown in **Fig. 9i** and **Fig. 10j**, respectively, while direct comparisons for each alcohol system are presented in **Fig. 11k–m**.

A clear evolution in morphology is observed with increasing alcohol chain length. The lateral dimensions of the perovskite flakes generally increase while their vertical dimensions decrease (**Fig. 12i**), indicating a progressive shift toward lateral growth relative to out-of-plane growth. Methanol, which contains a single carbon atom (C1), typically does not facilitate the formation of ultrathin flakes under these conditions (**Supplementary Fig. 1a**). Ultrathin flakes begin to form when ethanol (C2) is used, while increasing the chain length to propanol (C3) and butanol (C4) produces progressively larger and thinner flakes. The molecular structure of the alcohol also influences the resulting morphology. 1-Propanol, with a linear carbon chain, produces generally larger and thinner flakes than its branched isomer, 2-propanol (**Fig. 13b,c**). These observations indicate that both alcohol chain length and molecular structure influence the balance between lateral and out-of-plane crystal growth.

To further reduce flake aggregation and increase the population of laterally extended ultrathin flakes, ODA was introduced into the solvent system as a long-chain steric additive. The C18 alkyl chain of ODA provides substantial steric bulk, while its terminal amine group may interact with organic and ionic species within the crystallization environment. ODA is therefore expected to hinder the close approach and restacking of neighbouring flakes during growth, promoting better spatial isolation. The paired AFM images in **Fig. 14f–h** show reduced aggregation in the presence of ODA compared with the corresponding ODA-free systems. The optical microscopy images of the ODA-containing samples are provided in **Supplementary Fig. 4**. The combined height–width distribution in **Fig. 15j**, together with the direct comparisons in **Fig. 16k–m**, further shows an increased population of thinner, laterally extended flakes following ODA incorporation. The combined use of longer-chain alcohol cosolvents and ODA therefore shifts the morphology toward larger lateral dimensions and reduced thickness, increasing the yield of well-isolated ultrathin flakes, predominantly monolayers.

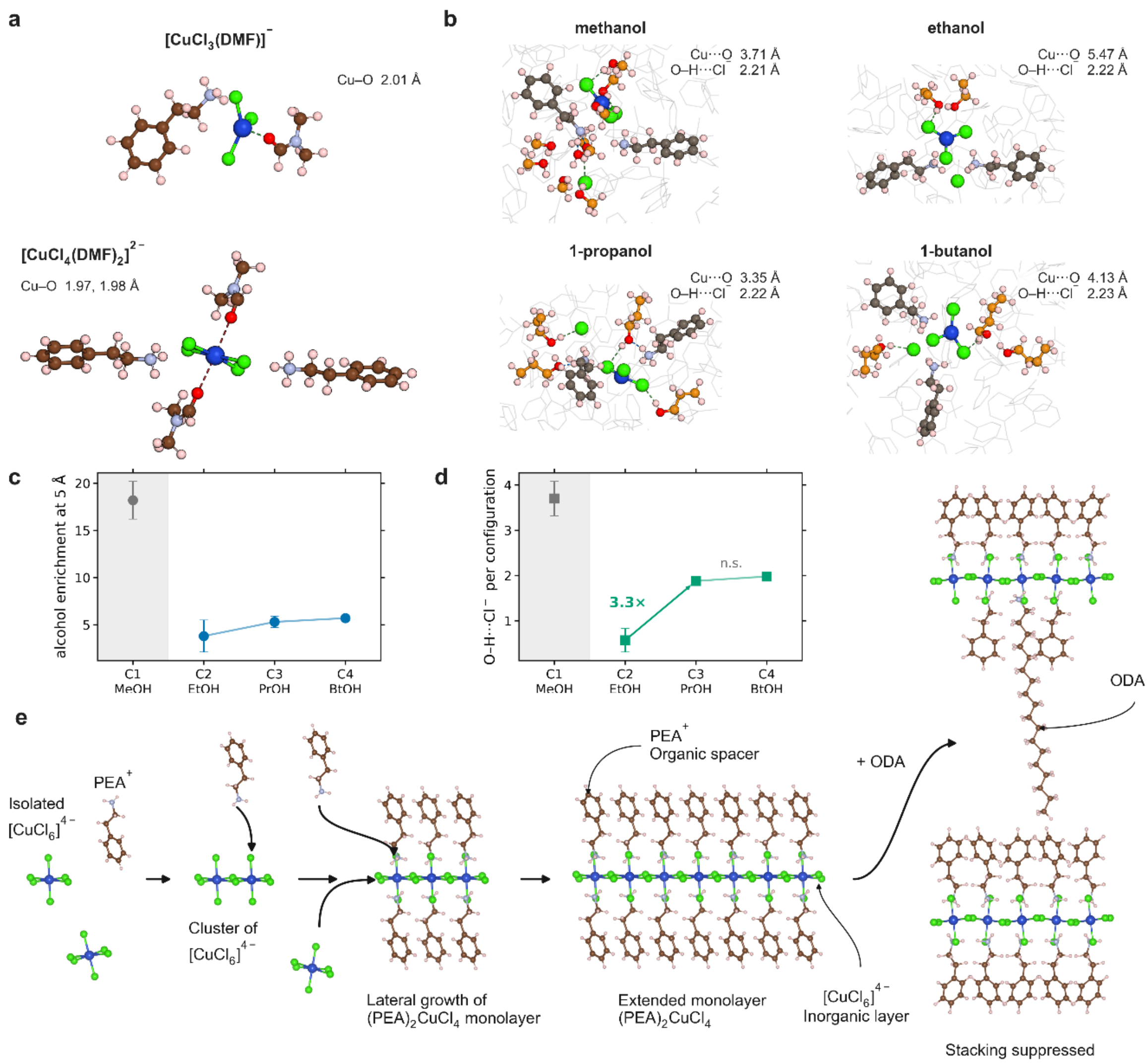

**Fig. 17: Computational insights: precursor solvation and the lateral-growth mechanism of ultrathin $(PEA)_2CuCl_4$. (a)** Quantum-chemically optimized structures of representative chloride/DMF-coordinated $Cu^{2+}$ precursor complexes, illustrating the primary coordination of DMF to the Cu centre and the corresponding Cu–O distances. Cluster geometries were optimized at the r$^2$SCAN-3c level. **(b)** Representative condensed-phase molecular-dynamics configurations of the $[CuCl_4]^{2-}$/$PEA^+$ precursor in chlorobenzene containing methanol, ethanol, 1-propanol, or 1-butanol. In every simulation the initial $[CuCl_4]^{2-}$ loses one chloride early in the trajectory, so the species sampled is $[CuCl_3]^-$ with a free chloride and two $PEA^+$. Insets show the closest Cu–O (alcohol) separation and the average number of O–H···$Cl^-$ hydrogen bonds per configuration. Mean closest Cu–O separations of ~3.4–5.5 Å for ethanol, 1-propanol and 1-butanol are substantially longer than the ~2.1 Å characteristic of direct Cu–O coordination, indicating that these alcohols remain outside the primary Cu coordination sphere and interact predominantly with the surrounding ionic precursor environment;

methanol transiently approaches to 2.1 Å. **(c)** Alcohol enrichment within 5 Å of the Cu centre, normalized to the bulk alcohol:chlorobenzene composition. **(d)** Average number of O–H···$Cl^-$ hydrogen bonds per configuration. Ethanol forms fewer hydrogen bonds to chloride than 1-propanol and 1-butanol, whereas the latter two are not statistically distinguishable (*n.s.*). Methanol, shown separately in the shaded region in **(c,d)**, exhibits anomalously strong precursor association and represents a distinct crystallization regime rather than part of the C2–C4 trend. Error bars represent standard errors corrected for the integrated autocorrelation time of each observable within a single trajectory per alcohol. **(e)** Schematic of the proposed growth pathway, progressing from isolated Cu–Cl precursor units through precursor association and nucleation to lateral propagation of the $(PEA)_2CuCl_4$ inorganic framework between $PEA^+$ organic layers. Moderation of precursor association and nucleation, followed by continued precursor transport to exposed crystal edges, favours lateral monolayer growth over out-of-plane thickening.

The formation of ultrathin $(PEA)_2CuCl_4$ crystals can be understood as a sequential process involving precursor desolvation, nucleation, and lateral crystal growth. In the initial DMF stock solution, $Cu^{2+}$ is strongly coordinated by the carbonyl oxygen of DMF. In the presence of chloride from $CuCl_2$ and PEACl, it is expected to exist as a dynamic distribution of chloride/DMF-coordinated species, represented generally as $[CuCl_x(DMF)_y]^{2-x}$ [19,20]. Quantum-chemical calculations support direct Cu–O coordination by DMF, with an optimized Cu–O distance of approximately 2.0 Å for $[CuCl_3(DMF)]^-$ (**Fig. 18a**). The strong coordinating ability and high polarity of DMF stabilize these ionic precursor species and suppress premature association. When 10 µL of this stock solution is injected into the much larger chlorobenzene-alcohol binary solvent mixture (chlorobenzene: Alcohol = 3200:400 µL), the local DMF concentration decreases and the Cu–Cl–DMF solvation shell progressively reorganizes. Because CB is weakly coordinating and poorly stabilizes ionic Cu–Cl/PEA species, precursor solubility decreases, driving association, supersaturation, and nucleation [21]. We propose that the alcohol moderates this otherwise abrupt change by reorganizing the secondary solvation environment around the ionic precursor, primarily through hydrogen-bonding interactions with chloride and the ammonium groups of $PEA^+$ rather than by directly coordinating $Cu^{2+}$ [22,23]. Representative molecular-dynamics configurations for methanol, ethanol, 1-propanol, and 1-butanol are shown in **Fig. 19b**. In the molecular-dynamics simulations, which contain no DMF, the mean closest Cu–O separations for ethanol, 1-propanol and 1-butanol are approximately 3.4–5.5 Å, substantially longer than the ~2.1 Å expected for a Cu–O dative bond, whereas methanol transiently approaches to 2.1

Å. The local association also increases markedly for the longer-chain alcohols: the average number of O–H···Cl⁻ contacts increases from 0.57 for ethanol to 1.88 for 1-propanol and 1.98 for 1-butanol **(Fig. 20d)**, while the corresponding precursor-enrichment factor increases from 3.8 ± 1.7 to 5.3 ± 0.6 and 5.7 ± 0.3 **(Fig. 21c)**, respectively. Quantum-chemical calculations indicate that exchange of DMF for alcohol at Cu is nearly thermoneutral at standard state (+0.04 to +0.08 eV in chlorobenzene; Supplementary Table 3), so the calculations do not resolve the Cu coordination state in the growth solution. The robust computational observation is alcohol–chloride hydrogen bonding in the secondary solvation environment.

The resulting supersaturation kinetics determine nucleation density and crystal morphology. In CB alone, the abrupt loss of precursor solvation generates high supersaturation, lowering the nucleation barrier and promoting rapid formation of a larger population of stable nuclei [24]. Under these conditions, precursor is distributed among many crystallisation centres and continued nucleation, and multilayer accumulation compete strongly with lateral propagation, resulting predominantly in thick, bulk-like layered crystallites. Addition of an alcohol is proposed to moderate the increase in supersaturation and thereby to reduce the nucleation rate, allowing more precursor to feed existing nuclei rather than continuously generating new ones. The greater precursor association observed for 1-propanol and 1-butanol (**Fig. 22c,d)** is consistent with a more strongly reorganized secondary solvation environment. We propose that this favours formation of ultrathin crystals and allows the intrinsic anisotropy of $(PEA)_2CuCl_4$ to be expressed more effectively through in-plane growth. Importantly, the increase in association is not simply caused by stronger individual alcohol–chloride hydrogen bonds. The calculated intrinsic ROH···Cl⁻ interaction changes by only ~30 meV across the C1–C4 alcohol series, whereas the number of alcohol–chloride contacts increases by more than a factor of three between ethanol and the longer alcohols **(Fig. 23d)**. This suggests that the difference between ethanol and the longer alcohols more likely reflects how the alcohol molecules are accommodated around the precursor than individual hydrogen-bond strength.

Once an ultrathin layer has formed, however, the growth mechanism changes from nucleation control to precursor transport. In the n = 1 Ruddlesden–Popper structure, the ammonium end of $PEA^+$ is bound adjacent to the inorganic Cu–Cl sheet, whereas the phenethyl groups extend outward and terminate the basal surfaces. Consequently, alcohol-mediated interactions are most relevant before and during nucleation, or at incomplete layers and exposed crystal edges, rather than on a fully formed PEA-terminated basal surface. Further lateral enlargement therefore requires residual Cu–Cl/PEA precursor species to remain mobile in the surrounding liquid, diffuse toward exposed crystal edges, and incorporate into the advancing lattice. This proposed progression from precursor association to cluster formation and preferential lateral propagation of the $(PEA)_2CuCl_4$ monolayer is illustrated schematically in **Fig. 24e**.

This makes the lifetime of the wet growth environment particularly important during annealing. Notably, although 1-butanol experimentally produces larger flakes than 1-propanol, the two alcohols are not statistically distinguishable in the calculated solvation metrics **(Fig. 25c,d)**. This indicates that solvation effects alone cannot explain the additional increase in lateral crystal size from 1-propanol to 1-butanol. Ethanol can moderate the initial nucleation process and thereby produce thin flakes, but its lower viscosity and higher volatility shorten the lifetime of the liquid phase. The time available for continued precursor transport to the crystal edges is therefore limited, resulting in thin but laterally smaller flakes. In contrast, 1-butanol is more viscous and less volatile, so the liquid precursor environment persists for longer during annealing. Similar perovskite-processing studies have shown slower removal of n-butanol than ethanol and a correspondingly prolonged solution-mediated recrystallisation period [25], while solvent-evaporation kinetics are known to influence lateral growth in 2D Ruddlesden–Popper perovskites[26]. Increasing the alcohol chain length beyond C4 does not improve the monolayer formation (**Supplementary Fig. 5**). This limitation arises from the differential evaporation rates within the ternary solvent.

The overall morphology therefore reflects two consecutive solvent-controlled stages, as summarized schematically in **Fig. 26e**. First, the alcohol regulates the transition from the DMF-solvated precursor to the nucleated crystal, suppressing the rapid precipitation and multilayer accumulation observed in CB alone. Second, after a thin nucleus has formed, the persistence of the liquid phase during annealing determines how long precursor species can continue feeding the exposed crystal edges. This is consistent with the experimental progression from thick crystallites in CB, to smaller thin flakes in ethanol, and finally to large laterally extended thin flakes in 1-butanol. We propose that 1-butanol provides a favourable balance between sufficient driving force for crystallisation, moderated precursor solvation reorganisation and nucleation, and slow solvent removal to maintain a mobile precursor reservoir during annealing.

**Steric engineering: Universal synthesis method**

Having established that steric hindrance and secondary-solvation tuning govern the balance between nucleation density and lateral precursor transport, we reasoned that this physical mechanism should be inherently system-agnostic. Because the approach relies on regulating liquid-phase precursor association rather than specific substrate epitaxy, it offers a general knob for controlling the crystallization kinetics across the broad chemical family of $A_2MX_4$ hybrid perovskites. To demonstrate the universality of this strategy, we systematically extended the steric-engineering platform across all three functional sublattices: the organic spacer cation (A), the divalent metal center (M), and the halide anion (X) (**Error! Reference source not found.**Error! Reference source not found.**a**). In the following subsections, we explore how tuning the synthesis parameters to accommodate the distinct steric bulk, electronic interactions, and solubility profiles of these building blocks yields laterally extended, high-quality single- and few-layer crystals across a diverse compositional landscape.

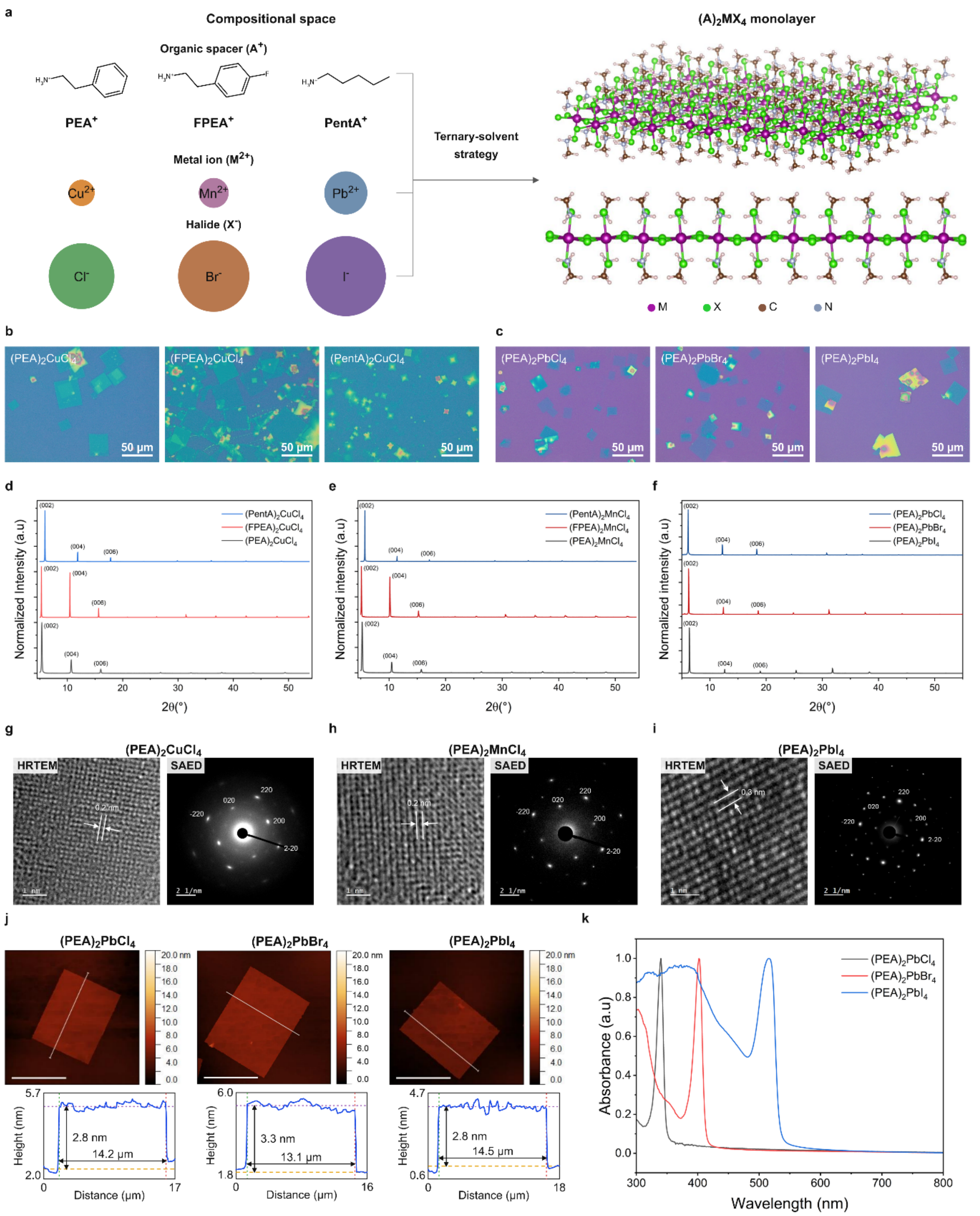

**Fig. 27: Universal ternary-solvent strategy for the growth of ultrathin $A_2MX_4$ 2D hybrid perovskites.**

**(a)** Schematic illustrating the broad compositional space accessible using the ternary-solvent growth strategy, comprising the organic spacer cations phenethylammonium ($PEA^+$), 4-fluorophenethylammonium ($FPEA^+$), and n-pentylammonium ($PentA^+$); divalent metal cations $Cu^{2+}$, $Mn^{2+}$, and $Pb^{2+}$; and halide anions $Cl^-$, $Br^-$, and $I^-$. Top- and side-view structural models illustrate the monolayer $A_2MX_4$ 2D hybrid perovskite structure. **(b)** Optical microscopy images of $(PEA)_2CuCl_4$, $(FPEA)_2CuCl_4$, and $(PentA)_2CuCl_4$, showing large-area ultrathin flakes and demonstrating the applicability of the growth strategy to different organic spacer cations. **(c)** Optical microscopy images of large area $(PEA)_2PbCl_4$, $(PEA)_2PbBr_4$, and $(PEA)_2PbI_4$ flakes, demonstrating extension of the strategy across the halide series. **(d–f)** XRD patterns of **(d)** $A_2CuCl_4$ based hybrid perovskites with different organic spacer cations, **(e)** $A_2MnCl_4$ based hybrid perovskites with different organic spacer cations, and **(f)** the $(PEA)_2PbX_4$ series (X = Cl, Br, and I), measured using Co Kα radiation (λ = 1.79 Å). The characteristic low-angle (00l) reflections confirm the layered crystal structure and reveal composition-dependent variations in the out-of-plane periodicity. **(g–i)** HRTEM images and corresponding SAED patterns of **(g)** $(PEA)_2CuCl_4$, **(h)** $(PEA)_2MnCl_4$, and **(i)** $(PEA)_2PbI_4$. Lattice spacings of ≈0.2 nm are observed for $(PEA)_2CuCl_4$ and $(PEA)_2MnCl_4$ and ≈0.3 nm for $(PEA)_2PbI_4$, while the discrete SAED spots confirm the crystalline nature of the flakes. **(j)** AFM topography images and corresponding height profiles of monolayer $(PEA)_2PbCl_4$, $(PEA)_2PbBr_4$, and $(PEA)_2PbI_4$ flakes, showing apparent heights of ≈2.8, 3.3, and 2.8 nm, respectively. **(k)** UV–vis absorbance spectra of $(PEA)_2PbCl_4$, $(PEA)_2PbBr_4$, and $(PEA)_2PbI_4$, showing prominent excitonic absorption features at ≈340, 403, and 520 nm, respectively, corresponding to transition energies of ≈3.65, 3.08, and 2.38 eV and demonstrating a progressive redshift from Cl to Br to I.

**Organic cation**

The choice of organic spacer cations plays a critical role in directing the crystallization kinetics and morphology of hybrid perovskite 2D layers.

| Organic cation | Mn | | Cu | |
|---|---|---|---|---|
| | 1st 2θ peak position (°) | d-spacing (Å) | 1st 2θ peak position (°) | d-spacing (Å) |
| PEA | 5.25 | 19.6 | 5.37 | 19.1 |
| F-PEA | 5.12 | 20.0 | 5.29 | 19.4 |
| PentA | 5.72 | 17.9 | 5.93 | 17.3 |

**Table 1: Low angle XRD peak position and calculated d-spacing for Cu and Mn 2D hybrid perovskite with different organic spacer cations. The XRD data was collected using Co Kα radiation (λ = 1.79 Å).**

**Error! Reference source not found.**Error! Reference source not found.**a** shows the three different spacers investigated: phenethylammonium ($PEA^+$), 4-fluorophenethylammonium ($F\text{-}PEA^+$), and n-pentylammonium ($PentA^+$).

Copper ($Cu^{2+}$) and manganese ($Mn^{2+}$) based 2D hybrid perovskite ultrathin flakes were synthesized using these spacers in a ternary solvent system containing 1-butanol as the alcohol cosolvent. Optical microscopy images of the resulting copper-based flakes are shown in **Error! Reference source not found.b**, and those of the manganese-based flakes are shown in **Supplementary Figs. 9-11** for $PEA^+$, $F\text{-}PEA^+$ and $PentA^+$, respectively. The optical microscope images of ultrathin layers formed with different alcohol cosolvents are provided in **Supplementary Figs. 7-11**. Steric engineering enables large-area (>1000 $\mu m^2$) monolayer and few layer flakes across all three organic spacer cations.

**Error! Reference source not found.d,e** show the XRD patterns for copper ($Cu^{2+}$) and manganese ($Mn^{2+}$) based 2D perovskites, respectively, using these three different organic spacers. The characteristic low-angle (00l) reflections confirm the periodic layered structures of these crystals, with layer separation of 19.1 Å for $(PEA)_2CuCl_4$ and 19.6 Å for $(PEA)_2MnCl_4$, matching with published data [27,28].

Although $PEA^+$ and $F\text{-}PEA^+$ are structurally similar in size, the incorporation of a fluorine atom in the para position of the $F\text{-}PEA^+$ cation introduces significant electronic effects [29]. The strong electronegativity of fluorine enhances the quadrupole moment of the phenyl ring, fostering stronger $\pi$–$\pi$ interactions between adjacent organic layers. This results in a more co-facially aligned and tightly packed organic sublattice[30]. Experimentally, formation of well-defined ultrathin FPEA-based crystals required a slightly higher alcohol fraction than the corresponding PEA-based system, indicating that modification of the organic spacer alters the solvent conditions required for controlled crystallization.

$PentA^+$ presents a distinct case because it is an aliphatic spacer and therefore lacks the aromatic $\pi$–$\pi$ interactions available to $PEA^+$ and $FPEA^+$. Consequently, its interaction with the evolving inorganic framework and solvent environment differs substantially from those of the aromatic spacers. We found that PentA-based precursors were less tolerant of high alcohol fractions, and well-defined 2D layers were obtained using a lower alcohol content together with elevated annealing temperatures. These differences in crystallization conditions are reflected in the resulting morphology, with PentA-based crystals generally exhibiting different lateral dimensions and thickness distributions compared with PEA- and FPEA-based crystals (**Supplementary Fig. 7-11**).

**Metal cation**

We synthesized monolayers in the $(PEA)_2MX_4$ (M: $Cu^{2+}$, $Mn^{2+}$, $Pb^{2+}$) system to demonstrate the universal nature of the steric engineering approach. Error! Reference source not found.**g-i** show the high-resolution TEM images and the SAED patterns of $(PEA)_2CuCl_4$, $(PEA)_2MnCl_4$, and $(PEA)_2PbI_4$, respectively. Large-area perovskite crystals are observed in the low magnification TEM images. We used thick crystals for imaging to mitigate the effect of the electron beam damage on the image quality[31]. All perovskite layers are highly crystalline, as evidenced by the high-resolution TEM images. The SAED patterns confirm the crystalline nature and long-range structural order of the perovskite samples. The SAED patterns match with the simulated electron diffraction patterns shown in **Supplementary Fig. 6**. The diffraction pattern confirms a preferential crystallographic orientation with layered inorganic framework (ab-plane) of the perovskite lattice parallel to the substrate. The lattice spacing, extracted from the diffraction pattern are 3.7 Å, 3.6 Å and 2.6 Å for $(PEA)_2CuCl_4$, 3.6 Å, 3.6 Å and 2.5 Å for $(PEA)_2MnCl_4$ and 4.3 Å, 4.3 Å and 3.1 Å for $(PEA)_2PbI_4$ respectively, in the (020), (200) and (220) directions. The crystal structure and associated parameters extracted from structural characterization match with published literature for $(PEA)_2CuCl_4$, $(PEA)_2MnCl_4$ and $(PEA)_2PbI_4$[27,28,32].

**Halide anion**

The compositional versatility of the growth strategy was further investigated across the $(PEA)_2PbX_4$ halide series (X = $Cl^-$, $Br^-$, and $I^-$). Error! Reference source not found.**c** shows the optical microscope images of $(PEA)_2PbCl_4$, $(PEA)_2PbBr_4$ and $(PEA)_2PbI_4$ synthesized using ternary solvent method. XRD patterns in **Fig. 4f** exhibit the characteristic low-angle (00l) reflections confirms the layered structure of the perovskites synthesized.

AFM topography and corresponding height profiles in **Fig. 4j** further confirm the formation of monolayer-scale crystals, with apparent step heights of approximately 2.8, 3.3, and 2.8 nm for $(PEA)_2PbCl_4$, $(PEA)_2PbBr_4$, and $(PEA)_2PbI_4$, respectively.

UV–visible absorption spectroscopy (Error! Reference source not found.**k**) reveal the tunable excitonic character of the 2D lead hybrid perovskite series $(PEA)_2PbX_4$ (X = $Cl^-$, $Br^-$, $I^-$). The absorption spectra exhibit a redshift as the halide changes from chloride to iodide, indicating bandgap narrowing. The origin of this trend lies in the electronic structure: the valence-band maximum has substantial antibonding Pb 6s–Xp character, whereas the conduction-band minimum is dominated primarily by Pb 6p states. As the halide is varied from Cl to Br to I, the increasing energy of the halide p orbitals raises the valence-band edge and reduces the separation between the valence and conduction bands, giving rise to the observed redshift[33,34].

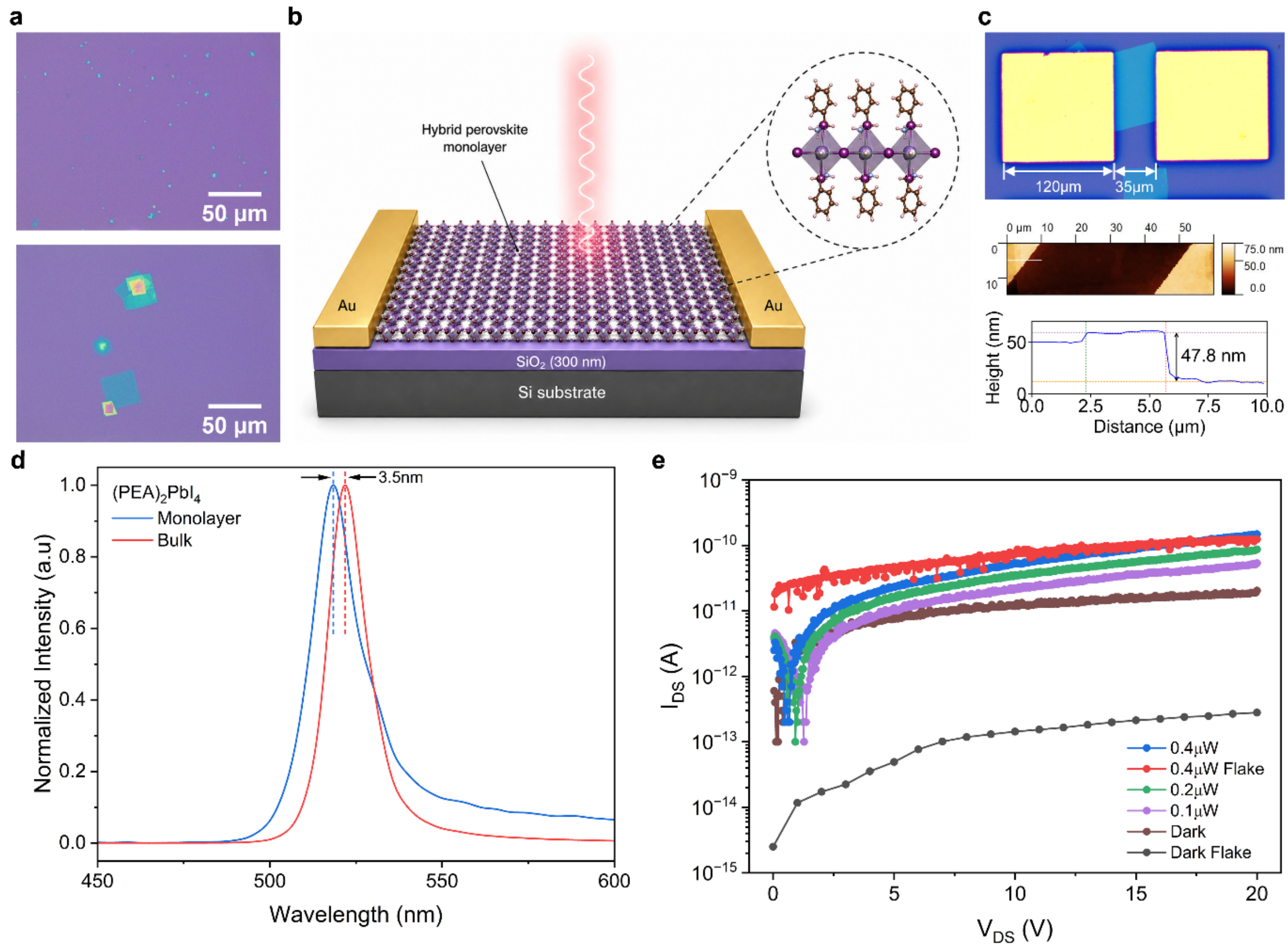

**Fig. 28: Ultra-thin $(PEA)_2PbI_4$ planar device platform and optoelectronic transport characteristics.**

**(a)** Optical microscopy images of $(PEA)_2PbI_4$ flakes obtained using the conventional ternary-solvent method (top) and the sterically engineered ternary-solvent method (bottom), demonstrating the substantial increase in lateral dimensions achieved through steric engineering. **(b)** Schematic illustration of the planar two terminal device architecture, in which an ultrathin $(PEA)_2PbI_4$ flake is contacted via Au electrodes on a $SiO_2$ (300 nm)/Si substrate. **(c)** Optical microscopy image of the fabricated device with an electrode width of ≈120 μm and channel length of ≈35 μm. AFM topography across the device and corresponding height profile, showing an Au electrode thickness of ≈47.8 nm. d) Normalized photoluminescence (PL) spectra of monolayer and bulk $(PEA)_2PbI_4$, revealing an ≈3.5 nm blueshift of the monolayer emission relative to the bulk crystal. **(e)** Output characteristics $I_{DS}$ vs $V_{DS}$ of $(PEA)_2PbI_4$ devices measured in the dark and under 532 nm illumination at incident powers of 0.1, 0.2, and 0.4 μW. The ultrathin-flake device exhibits a substantially reduced dark current, down to the $10^{-15}$–$10^{-13}$ A range (yielding femtowatt-to-picowatt static power consumption), compared with ≈$10^{-12}$–$10^{-11}$ A for the drop-cast device, while maintaining a high light-to-dark signal contrast ratio of $4.3 \times 10^2$.

The sterically engineered large-area crystals address a fundamental bottleneck in 2D hybrid perovskite device fabrication. Standard lithographic patterning relies on organic resists, developers, and strippers that cause irreversible chemical degradation and phase decomposition in metal-halide lattices. Previously reported, conventional ternary-solvent approach produces predominantly small, sparsely distributed flakes (**Fig. 28a**, top), that cannot be contacted directly without the damaging post-processing steps. The sterically engineered strategy yields significantly larger and laterally extended crystals (**Fig. 28a**, bottom), providing dimensions suitable for non-destructive electrical contact formation via physical shadow masking. To evaluate the intrinsic optoelectronic transport and low-power capabilities of these un-degraded channels, we integrated individual ultrathin $(PEA)_2PbI_4$ crystals into planar optoelectronic device platforms.

A planar optoelectronic device was fabricated using an individual ultrathin $(PEA)_2PbI_4$ flake as the active channel between Au electrodes on a 300 nm $SiO_2$/Si substrate (**Fig. 28b**). The device has an electrode width of ≈120 µm and a channel length of ≈35 µm. AFM mapping across the electrode region and the corresponding line profile confirm an Au electrode thickness of ≈48 nm and perovskite layer thickness of ≈10 nm (**Fig. 28c**). The reduction in crystal thickness is accompanied by measurable changes in the optical response. As shown in **Fig. 28d**, the monolayer exhibits a ~3.5 nm blueshift in the PL peak relative to the bulk crystal, indicating a modest modification of the excitonic transition energy in the ultrathin limit. Such a shift is consistent with changes in dielectric screening, structural relaxation, and the local surface/interface environment as the number of stacked perovskite layers is reduced.

The electrical and photo-switching characteristics highlight the decisive advantage of using an isolated, shadow-masked ultrathin crystal as a low-power device channel (**Fig. 28e**). Across the measured bias range, the drop-cast reference device exhibits a high dark current baseline on the order of $10^{-12}$–$10^{-11}$ A, whereas the ultrathin crystal channel displays an ultra-low, suppressed dark current baseline down to the $10^{-15}$–$10^{-13}$ A range. At an operating bias of $V_{DS}$ = 5 V, the static power consumption of the isolated ultrathin channel in standby mode is $P_{static}$ is ~ 0.2 pW (200 fW), increasing to only 5.6 pW at $V_{DS}$ = 20V. This represents a 250 times reduction in static power dissipation compared to drop-cast control devices (~ 50 pW at 5 V), establishing a low-power device baseline enabled by damage-free contact integration.

Importantly, this dramatic suppression of dark current and static power is achieved without compromising photo-driven carrier generation: under 532 nm illumination at 0.4 µW, both drop-cast and ultrathin flake devices yield comparable photocurrents of ~$1.4 \times 10^{-10}$ and ~$1.2 \times 10^{-10}$ A, respectively at $V_{DS}$ = 20V. Consequently, the light-to-dark current ratio increases from only ~8.1 for the drop-cast device to ~$4.3 \times 10^2$ for the ultrathin isolated flake. The femtoampere-level dark baseline underscores the high crystal purity, low defect density, and pristine contact interfaces achieved via physical shadow masking. Eliminating dark carrier injection and parasitic leakage paths provides a crucial noise-floor benchmark for single-layer 2D optoelectronics, demonstrating that scalable steric engineering can yield device-grade, low-noise active channels without requiring post-synthetic chemical processing.

**Conclusion**

In summary, we have developed a universal steric-engineering ternary-solvent strategy for the solution-phase growth of laterally extended, high-quality $A_2MX_4$ hybrid perovskite monolayers and few-layer crystals with surface areas exceeding 1000 µm$^2$. By synergistically pairing long-chain alcohol cosolvents with an amine steric additive (octadecylamine), out-of-plane stacking is suppressed while spatial isolation between growing crystals is enhanced. Quantum-chemical calculations and molecular dynamics simulations suggest that longer-chain alcohols reorganize the secondary solvation shell around precursor complexes via O-H···$Cl^-$ hydrogen bonding, while reduced solvent volatility prolongs edge-mediated precursor transport during annealing to favor anisotropic in-plane propagation. Crucially, this strategy demonstrates broad synthetic universality across diverse organic spacer cations ($PEA^+$, F-$PEA^+$, $PentA^+$), metal centers ($Cu^{2+}$, $Mn^{2+}$, $Pb^{2+}$), and halide anions ($Cl^-$, $Br^-$, $I^-$). Beyond fundamental synthesis, the large lateral dimensions circumvent long-standing lithographic and solvent-degradation bottlenecks, enabling direct, non-destructive device fabrication via physical shadow masking. Planar optoelectronic devices from isolated ~10 nm thick $(PEA)_2PbI_4$ active channels exhibit ultra-low dark currents down to $10^{-15}$ - $10^{-13}$ A, operating in an exceptional low-power regime with static standby power dissipation down to ~ 0.2 pW at 5 V, a 250-fold reduction compared to drop-cast control devices, while maintaining a light-to-dark signal contrast ratio of $4.3 \times 10^2$. This work establishes steric and solvation control as a versatile paradigm for producing compositionally diverse ultrathin 2D perovskites, paving the way for exploring unscreened low-dimensional physics and realizing high-performance single-crystal 2D optoelectronic devices.

## References and Notes

1. Kojima, A., Teshima, K., Shirai, Y. & Miyasaka, T. Organometal Halide Perovskites as Visible-Light Sensitizers for Photovoltaic Cells. *Journal of the American Chemical Society* **131**, 6050–6051 (2009).
2. Jeon, N. J. *et al.* Solvent engineering for high-performance inorganic–organic hybrid perovskite solar cells. *Nature Mater* **13**, 897–903 (2014).
3. Hu, J., Yan, L. & You, W. Two-Dimensional Organic–Inorganic Hybrid Perovskites: A New Platform for Optoelectronic Applications. *Advanced Materials* **30**, 1802041 (2018).
4. Grancini, G. & Nazeeruddin, M. K. Dimensional tailoring of hybrid perovskites for photovoltaics. *Nature Reviews Materials* vol. 4 4–22 (2019).
5. Mao, L., Stoumpos, C. C. & Kanatzidis, M. G. Two-Dimensional Hybrid Halide Perovskites: Principles and Promises. *J. Am. Chem. Soc.* **141**, 1171–1190 (2019).
6. Li, X., Hoffman, J. M. & Kanatzidis, M. G. The 2D Halide Perovskite Rulebook: How the Spacer Influences Everything from the Structure to Optoelectronic Device Efficiency. *Chem. Rev.* **121**, 2230–2291 (2021).
7. Katan, C., Mercier, N. & Even, J. Quantum and Dielectric Confinement Effects in Lower-Dimensional Hybrid Perovskite Semiconductors. *Chemical Reviews* **119**, 3140–3192 (2019).
8. Blancon, J.-C. *et al.* Scaling law for excitons in 2D perovskite quantum wells. *Nature Communications* **9**, (2018).
9. Even, J., Pedesseau, L. & Katan, C. Understanding Quantum Confinement of Charge Carriers in Layered 2D Hybrid Perovskites. *ChemPhysChem* **15**, 3733–3741 (2014).
10. Yaffe, O. *et al.* Excitons in ultrathin organic-inorganic perovskite crystals. *Phys. Rev. B* **92**, 045414 (2015).
11. Niu, W., Eiden, A., Vijaya Prakash, G. & Baumberg, J. J. Exfoliation of self-assembled 2D organic-inorganic perovskite semiconductors. *Applied Physics Letters* **104**, (2014).
12. Leng, K. *et al.* Molecularly thin two-dimensional hybrid perovskites with tunable optoelectronic properties due to reversible surface relaxation. *Nature Materials* **17**, 908–914 (2018).
13. Niu, T. *et al.* Reduced-Dimensional Perovskite Enabled by Organic Diamine for Efficient Photovoltaics. *The Journal of Physical Chemistry Letters* **10**, 2349–2356 (2019).

14. Dou, L. *et al.* Atomically thin two-dimensional organic-inorganic hybrid perovskites. *Science* **349**, 1518–1521 (2015).

15. Leng, K., Fu, W., Liu, Y., Chhowalla, M. & Loh, K. P. From bulk to molecularly thin hybrid perovskites. *Nature Reviews Materials* **5**, 482–500 (2020).

16. Chen, J. *et al.* A Ternary Solvent Method for Large-Sized Two-Dimensional Perovskites. *Angewandte Chemie* **129**, 2430–2434 (2017).

17. Willett, R. D. Structures of the antiferrodistortive layer perovskites bis(phenethylammonium) tetrahalocuprate(II), halo = Cl−, Br−. *Acta Cryst C* **46**, 565–568 (1990).

18. Nemes-Incze, P., Osváth, Z., Kamarás, K. & Biró, L. P. Anomalies in thickness measurements of graphene and few layer graphite crystals by tapping mode atomic force microscopy. *Carbon* **46**, 1435–1442 (2008).

19. Elleb, M., Meullemeestre, J., Schwing-Weill, M. J. & Vierling, F. Stability, electronic spectra, and structure of the copper(II) chloride complexes in N,N-dimethylformamide. *Inorg. Chem.* **19**, 2699–2704 (1980).

20. Yilmaz, V. T. & Topcu, Y. Preparation, characterization and thermal reactivity of *N*,*N*-dimethylformamide complexes of some transition metal chlorides. *Thermochimica Acta* **307**, 143–147 (1997).

21. Bautista-Quijano, J. R., Telschow, O., Paulus, F. & Vaynzof, Y. Solvent–antisolvent interactions in metal halide perovskites. *Chem. Commun.* **59**, 10588–10603 (2023).

22. Ichihashi, M., Wakita, H. & Masuda, I. X-Ray Diffraction Study on the Coordination Structure of Chlorocopper(II) Complexes in Highly Concentrated Methanolic CuCl2 Solution. *bull. Chem. Soc. Jpn.* **56**, 3761–3764 (1983).

23. Fridgen, T. D., McMahon, T. B., Maître, P. & Lemaire, J. Experimental infrared spectra of Cl−(ROH) (R = H, CH3, CH3CH2) complexes in the gas-phase. *Phys. Chem. Chem. Phys.* **8**, 2483–2490 (2006).

24. Sun, J. *et al.* Mechanistic insights and optimization strategies for perovskite single-crystal thin film growth. *Chem. Sci.* **16**, 6188–6202 (2025).

25. Zheng, X. *et al.* Solvent engineering for scalable fabrication of perovskite/silicon tandem solar cells in air. *Nature Communications* **15**, 4907 (2024).

26. Dai, Q. *et al.* Regulating Radial Morphology in Hot-Casting Two-Dimensional Ruddlesden–Popper Perovskite Film Growth for High-Efficient Photovoltaics. *ACS Appl. Energy Mater.* **6**, 1585–1594 (2023).

27. Polyakov, A. O. *et al.* Coexisting Ferromagnetic and Ferroelectric Order in a CuCl4-based Organic–Inorganic Hybrid. *Chemistry of Materials* **24**, 133–139 (2011).

28. Kamminga, M. E., Hidayat, R., Baas, J., Blake, G. R. & Palstra, T. T. M. Out-of-plane polarization in a layered manganese chloride hybrid. *APL Mater.* **6**, 066106 (2018).

29. García-Benito, I. *et al.* Fluorination of Organic Spacer Impacts on the Structural and Optical Response of 2D Perovskites. *Front. Chem.* **7**, (2020).

30. Zhang, F. *et al.* Enhanced Charge Transport in 2D Perovskites via Fluorination of Organic Cation. *J. Am. Chem. Soc.* **141**, 5972–5979 (2019).

31. Du, H.-Q. *et al.* Transmission electron microscopy studies of organic–inorganic hybrid perovskites: Advances, challenges, and prospects. *Appl. Phys. Rev.* **10**, 021314 (2023).

32. Straus, D. B. *et al.* Longer Cations Increase Energetic Disorder in Excitonic 2D Hybrid Perovskites. *J. Phys. Chem. Lett.* **10**, 1198–1205 (2019).

33. Walsh, A. Principles of Chemical Bonding and Band Gap Engineering in Hybrid Organic–Inorganic Halide Perovskites. *J. Phys. Chem. C* **119**, 5755–5760 (2015).

34. Butler, K. T., Frost, J. M. & Walsh, A. Band alignment of the hybrid halide perovskites CH3NH3PbCl3, CH3NH3PbBr3 and CH3NH3PbI3. *Mater. Horiz.* **2**, 228–231 (2015).

**Acknowledgments:**

This work was performed in part at the Melbourne Centre for Nanofabrication (MCN) in the Victorian Node of the Australian National Fabrication Facility (ANFF). This research was undertaken with the assistance of resources from the Monash University and National Computational Infrastructure (NCI Australia) allocation scheme. NCI is an NCRIS-enabled capability supported by the Australian Government.

**Funding:**

**Author contributions:**

Conceptualization: SK, SM

Methodology: SK, SD, SM

Investigation: SK, SD, SMS

Visualization: SM

Funding acquisition: SM

Project administration: SK, SM

Supervision: SM

Writing – original draft: SK

Writing – review & editing: SM, CM

**Diversity, equity, ethics, and inclusion [optional]:**

**Competing interests:**

Authors declare that they have no competing interests.

**Data and materials availability:**

**Supplementary Materials**

Supplementary Materials for

# Universal Synthesis of Large-Area Hybrid Perovskite Monolayers Enabling Lithography-Free, Low-Static-Power Optoelectronics

Shankeerthan Kalyanasundaram, Swarit Dwivedi, Sanje Mahasivam, Christopher R. McNeill, Sudha Mokkapati*

Corresponding author: Sudha.mokkapati@monash.edu

**The file includes:**

# 1. Materials and Methods

## 1.1 Materials

Copper (II) chloride ($CuCl_2$, anhydrous, 99.995%), manganese (II) chloride ($MnCl_2$, anhydrous, 99.995%), tin (II) chloride ($SnCl_2$, 99.99%), tin (II) bromide ($SnBr_2$, 99.99%), tin (II) iodide ($SnI_2$, 99.99%), lead chloride ($PbCl_2$, 99.99%), lead bromide ($PbBr_2$, 99.99%), lead iodide ($PbI_2$, 99.99%), octadecylamine (ODA), dimethylformamide (DMF, anhydrous, 99.8%), chlorobenzene (anhydrous, 99.8%), methanol (99.5%), ethanol (99.5%), 1-propanol (99.5%), 2-propanol (99.5%), and 1-butanol (99.5%) were obtained from Sigma-Aldrich. Phenethylammonium chloride [(PEA)Cl, anhydrous, 99.99%], 4-fluorophenethylammonium chloride [(F-PEA)Cl, anhydrous, 99.99%], n-pentylammonium chloride [(PentA)Cl, anhydrous, 99.99%] phenethylammonium bromide [(PEA)Br, anhydrous, 99.99%], and phenethylammonium iodide [(PEA)I, anhydrous, 99.99%] were purchased from Greatcell solar materials. All chemicals were used without further purification.

## 1.2 2D hybrid perovskite monolayer synthesis

### 1.2.1 Precursor solution preparation

The stock 2D hybrid perovskite precursor solutions were prepared by dissolving the corresponding metal halide and organic salt at a 1:2 molar ratio in dimethylformamide (DMF) to achieve a 0.1 M concentration. For copper- and manganese-based hybrid perovskites, $CuCl_2$ or $MnCl_2$ was mixed with the corresponding organic salt (A = phenethylammonium chloride [PEACl], 4-fluorophenethylammonium chloride [F-PEACl], or n-pentylammonium chloride [PentACl]). For the lead-based hybrid perovskites, $PbX_2$ and PEAX (X = $Cl^-$, $Br^-$, or $I^-$) were used. The mixtures were stirred at 60 °C for approximately 4 hours within a nitrogen-filled glovebox, followed by cooling to room temperature prior to use. The freshly prepared solutions were filtered through a 0.2 μm polytetrafluoroethylene (PTFE) syringe filter to remove undissolved particulates.

### 1.2.2 Hybrid perovskite monolayer synthesis

2D hybrid perovskite monolayers were synthesized by injecting the precursor solution dissolved in DMF into a binary solvent mixture of chlorobenzene and alcohol. Depending on the composition, methanol, ethanol, 1-propanol, 2-propanol, or 1-butanol was used as the alcohol component. The volume ratios of the resulting ternary mixture (chlorobenzene: alcohol: DMF) were optimized for each compound to obtain uniform ultrathin layers. The specific solvent ratios and annealing temperatures used for each composition are summarized in Supplementary Table 1.

**Supplementary Table 1.**

| **Composition** | **Chlorobenzene: Alcohol: DMF (µL)** | **Annealing temperature (°C)** |
|---|---|---|
| $(PEA)_2CuCl_4$ | 3200:400:10 | 50 °C |
| $(F\text{-}PEA)_2CuCl_4$ | 3100:500:10 | 60 °C |
| $(PentA)_2CuCl_4$ | 3308:300:2 | 75 °C |
| $(PEA)_2MnCl_4$ | 3100:500:10 | 50 °C |
| $(F\text{-}PEA)_2MnCl_4$ | 3000:600:10 | 50 °C |
| $(PentA)_2MnCl_4$ | 3300:300:10 | 65 °C |
| $(PEA)_2PbCl_4$ | 3000:600:10 | 50 °C |
| $(PEA)_2PbBr_4$ | 3000:600:30 | 55 °C |
| $(PEA)_2PbI_4$ | 3100:500:10 | 55 °C |

Following precursor injection, the resulting mixtures were allowed to settle undisturbed for 24 hours to enable complete growth of the ultrathin perovskite layers. For substrate deposition, 100 µL of the solution was drop-cast onto pre-cleaned $Si/SiO_2$ substrates (20 mm × 15 mm). The substrates were cleaned via sequential ultrasonication in ethanol and acetone for 10 minutes each, followed by a 15-minute UV-ozone treatment. Finally, the substrates were annealed for approximately 10 minutes at the temperatures listed in Supplementary Table 1 to ensure full solvent evaporation.

### 1.2.3 Synthesis of hybrid perovskite monolayers with additives

For additive-assisted synthesis, octadecylamine (ODA) was dissolved in different alcohols to form a $1 \times 10^{-4}$ M solution. The ODA solutions were added in varying volumes of 10 or 20 µL to the binary solvent mixture of chlorobenzene and alcohol. Thereafter, the $(PEA)_2CuCl_4$ precursor dissolved in DMF was injected into the binary solvent mixture containing the ODA additive and allowed to stabilize for 24 hours. Subsequently, 100 µL of the solution was drop-cast onto pre-cleaned $Si/SiO_2$ substrates using the same substrate-cleaning procedure described above. Finally, the substrates were annealed at 50 °C for approximately 10 minutes to ensure full solvent evaporation.

## 1.3 Characterization of 2D hybrid perovskite

### 1.3.1 Optical Microscopy

A Leica DM750 microscope was used to capture bright-field images of perovskite layers. Magnifications of 10×, 20×, and 50× were applied for imaging.

### 1.3.2 Atomic Force Microscopy

AFM measurements were performed on a Bruker Dimension Icon. Imaging and height profile analysis of perovskite monolayers and multilayers were carried out in non-contact tapping mode under ambient conditions, using a TESPA silicon probe (nominal resonance frequency ~320 kHz, spring constant ~42 N/m, tip radius <10 nm).

### 1.3.3 Photoluminescence Measurements

The photoluminescence spectra of hybrid perovskite layers were captured using a Witec confocal Raman microscope. A 355 nm ultrafast UV laser was used with varying integration times and power settings for the measurements.

### 1.3.4 Transmission Electron Microscopy

Transmission electron microscopy (TEM) imaging and selective area diffraction (SAED) analyses were conducted using a JEOL F200 equipped with a cold field emission gun (CFEG) operating at an accelerating voltage of 80 kV in TEM mode. Samples were drop-cast onto holey carbon-coated Cu-supported TEM grids for imaging in TEM and scanning transmission electron microscopy (STEM) modes. To minimize electron damage to the specimens, the TEM instrument employed a smaller aperture, spot size, and a current of 5 µA for this study. TEM images were captured using a Gatan One View camera.

### 1.3.5 XRD Measurements

Room-temperature XRD patterns of 2D hybrid perovskites were recorded using a Bruker D8 instrument with a cobalt source. The 2θ angles were varied from 5° to 55° during the measurements.

### 1.3.6 UV-Visible absorption spectroscopy

UV-Vis spectra were acquired using an Agilent Cary 365 spectrophotometer equipped with an integrating sphere. Measurements were performed in total transmittance mode from 200 to 900 nm. The samples were prepared as uniform coatings on fused quartz substrates to ensure transparency across the measured spectral range.

## 2 Molecular dynamics simulations

### 2.1 Computational methods

#### 2.1.1 Condensed-phase molecular dynamics

Simulation cells contained a single copper precursor unit ($[CuCl_4]^{2-}$ with two $PEA^+$ counter-cations, net charge zero by construction) solvated by an explicit chlorobenzene/alcohol mixture at the 1:7 alcohol:chlorobenzene molar ratio. Cells were cubic with edges of 36–37 Å and contained 2759–3023 atoms. Charge neutrality is required because the interatomic potential does not include Ewald summation, so a net-charged cell is unrepresentable. In every simulation, including the pure chlorobenzene control, one chloride dissociated from copper early in the trajectory; the species sampled is therefore $[CuCl_3]^-$ with a free chloride and two $PEA^+$ (Supplementary Table 2).

Forces were evaluated with the MACE-MPA-0 machine-learned interatomic potential together with D3(BJ) dispersion corrections using a 12 Å cutoff. Trajectories were propagated for 150 ps (methanol: 193 ps) in the NPT ensemble at 300 K and 1 atm with a 1 fs timestep. Four alcohols were simulated (methanol, ethanol, 1-propanol, 1-butanol) together with a pure chlorobenzene control. One trajectory was run per alcohol.

Analysis used every frame of the final 100 ps. Solvation shells were assigned using covalent-radius neighbour lists with a scaling factor of 1.25 and zero skin; the default skin setting misassigns O–H···Cl hydrogen bonds as covalent contacts. Uncertainties are standard errors corrected for the integrated autocorrelation time of each observable, so they reflect the number of statistically independent configurations rather than the number of frames.

#### 2.1.2 Alcohol association metric

Association of the alcohol with the ionic precursor was quantified as a size-independent local density ratio. One marker atom was assigned per molecule — the hydroxyl oxygen for the alcohols and the ring-bound chlorine for chlorobenzene — and the alcohol:chlorobenzene ratio within 5 Å of the copper centre was normalised by the bulk composition of the same cell. A value of unity denotes no preference. Using one marker per molecule removes the bias that would otherwise arise from the different molecular volumes of the four alcohols.

As a control, an analogous single-marker measure was evaluated in a separate analysis (final 50 ps) around randomly selected neutral chlorobenzene molecules in the same trajectories.

This tests whether any apparent enrichment is specific to the charged precursor or is an artefact of molecular packing.

#### 2.1.3 Cluster quantum chemistry

Ligand-exchange and hydrogen-bond energetics were computed for explicit clusters embedded in a continuum solvent. Geometries and harmonic frequencies were obtained at the $r^2$SCAN-3c level and single-point energies at ωB97X-D4/def2-TZVPP, with CPCM continuum solvation parameterised for chlorobenzene ($\varepsilon = 5.7$) and for DMF ($\varepsilon = 36.7$). Copper(II) is $d^9$ and was treated as a spin doublet throughout. Harmonic frequencies confirmed minima for all converged copper complexes. Free DMF retains one imaginary mode (196i $cm^{-1}$); because DMF appears in every exchange reaction, this shifts all values within a solvent column of Supplementary Table 3 by the same amount. Hydrogen-bond complexes with the organic cation used the complete phenethylammonium ion rather than a truncated analogue, so that the steric environment of the ammonium group is represented.

#### 2.1.4 Validation against density functional theory

The MACE-MPA-0 potential is trained predominantly on bulk inorganic crystals, whereas the present system is molecular, ionic and hydrogen-bonded. Ten independent condensed-phase configurations (≈400 atoms; $[CuCl_3]^-$ with $PEA^+$ and explicit solvent at liquid density, charge-neutral) taken from the MACE trajectories were therefore recomputed as single points with spin-polarised density functional theory (PBE with D3(BJ), 520 eV plane-wave cutoff, Γ-point sampling, spin-polarised with the total moment constrained to one unpaired electron per copper). These calculations quantify the accuracy of the machine-learned potential for this chemistry: the root-mean-square force error relative to DFT is 0.23 eV $Å^{-1}$ over all atoms and 0.29 eV $Å^{-1}$ for the Cu–Cl subset. Because the geometries were not re-optimised, they do not independently test the solvation structure.

### 2.2 Results: the precursor solvation environment

#### 2.2.1 Ethanol, 1-propanol and 1-butanol do not coordinate copper in the simulations

Ethanol, 1-propanol and 1-butanol remained outside the copper coordination sphere throughout. Mean closest Cu–O separations were 3.71 Å (methanol), 5.47 Å (ethanol), 3.35 Å (1-propanol) and 4.13 Å (1-butanol), with no systematic dependence on chain length. For comparison, a Cu–O dative bond in this system is ≈2.1 Å. Over the 1000 analysed configurations per alcohol, the minimum Cu–O separations were 3.27 Å (ethanol), 2.67 Å (1-propanol) and 2.94 Å (1-butanol). Methanol is the exception: it transiently approached to 2.08 Å, within bonding distance, and lay within 2.6 Å of copper in 23 of 1000 configurations. This distinction matters because the cluster calculations below place the alcohol at the metal by construction and therefore cannot test whether it goes there. In the explicit-solvent simulations, which contain no DMF, the C2–C4 alcohols do not. In these simulations the alcohol acts from the second solvation shell, through hydrogen bonding to chloride and to the ammonium groups of $PEA^+$, rather than by coordinating the metal centre.

### 2.2.2 Association with the ionic precursor

Supplementary Table 2 summarises the solvation metrics. The alcohols accumulate around the ionic precursor to an extent well beyond bulk composition, and the control analysis indicates that this is specific to the charged species: around neutral chlorobenzene molecules the analogous measure returns 0.99–1.10 for all four alcohols.

**Supplementary Table 2.** Solvation metrics from 150 ps molecular dynamics, averaged over the final 100 ps. Uncertainties are standard errors corrected for the integrated autocorrelation time. Control values are from the separate analysis described in Methods.

| Quantity | Methanol | Ethanol | 1-Propanol | 1-Butanol |
|---|---|---|---|---|
| Enrichment at 5 Å | 18.2 ± 2.0 | 3.8 ± 1.7 | 5.3 ± 0.6 | 5.7 ± 0.3 |
| Alcohol molecules within 5 Å of Cu | 3.12 | 0.60 | 1.02 | 0.99 |
| Chlorobenzene molecules within 5 Å of Cu | 0.41 | 1.06 | 1.32 | 0.92 |
| O–H···$Cl^-$ hydrogen bonds | 3.70 ± 0.38 | 0.57 ± 0.26 | 1.88 ± 0.07 | 1.98 ± 0.05 |
| of which to Cu-bound / free $Cl^-$ | 1.76 / 1.94 | 0.57 / 0.00 | 0.88 / 1.01 | 0.93 / 1.05 |
| Cl bound to Cu (mean) | 2.94 | 3.01 | 3.00 | 3.00 |
| Enrichment around neutral ClB (control) | 1.10 | 0.99 | 1.09 | 1.00 |
| Mean closest Cu–O (Å) | 3.71 | 5.47 | 3.35 | 4.13 |
| Autocorrelation time (ps) | 2.8 | 22.4 | 5.6 | 2.4 |

Ethanol differs from the two longer alcohols in its hydrogen bonding to chloride. It forms 0.57 ± 0.26 hydrogen bonds to chloride per configuration, against 1.88 ± 0.07 for 1-propanol and 1.98 ± 0.05 for 1-butanol — a factor of roughly three.

1-Propanol and 1-butanol, however, are not statistically distinguishable from one another. The difference in enrichment is 0.37 ± 0.71 (0.5σ) and in hydrogen-bond count 0.09 ± 0.08 (1.1σ). Neither approaches significance. The simulations therefore support a step in solvation behaviour between C2 and C3, but do not resolve a monotonic progression from C2 to C4.

## 2.3 Results: ligand-exchange and hydrogen-bond thermodynamics

### 2.3.1 Displacement of DMF from copper

The precursor enters the growth solution coordinated by DMF. The thermodynamics of its replacement by alcohol were evaluated for the reaction $[CuCl_3(DMF)]^-$ + ROH → $[CuCl_3(ROH)]^-$ + DMF (Supplementary Table 3).

**Supplementary Table 3.** Standard-state (1 M) free energy of DMF displacement by alcohol at the copper centre, at the r$^2$SCAN-3c level with CPCM solvation. Positive values indicate that the DMF-coordinated precursor is favoured. The 1-propanol value in chlorobenzene is omitted because the geometry optimisation of $[CuCl_3(\text{1-propanol})]^-$ in the chlorobenzene continuum did not converge.

| Alcohol | ΔG in chlorobenzene (eV) | ΔG in DMF (eV) |
|---|---|---|
| Methanol | +0.040 | +0.108 |
| Ethanol | +0.037 | +0.109 |
| 1-Propanol | not available | +0.171 |
| 1-Butanol | +0.075 | +0.140 |

Every value is positive but small: at standard state, alcohol and DMF compete almost equally for the fourth coordination site of $[CuCl_3]^-$, with a slight preference for DMF. Under the experimental conditions the alcohol is present in large excess over DMF ([ROH]/[DMF] ≈ 34–77, from 10 µL of DMF in 3.6 mL of solution), which contributes $-k^bT \ln([ROH]/[DMF]) \approx -0.09$ to $-0.11$ eV to the exchange free energy, so partial coordination of copper by the alcohol cannot be excluded by these calculations.

The differences between alcohols are small and method-dependent. In chlorobenzene the ethanol–butanol difference is 0.038 eV, approximately 1.5 $k^bT$ at room temperature, and is obtained from a continuum-solvent model. With ωB97X-D4 single-point energies and the same thermal corrections the ethanol and 1-butanol values become +0.023 and +0.060 eV, and with electronic energies alone their ordering reverses. In the DMF continuum the ordering is not monotonic in chain length (1-propanol +0.171 eV exceeds 1-butanol +0.140 eV), reinforcing that these differences are at the limit of what the method resolves. No chain-length trend in DMF displacement is therefore inferred.

### 2.3.2 Intrinsic alcohol–chloride hydrogen bond strength

To separate the strength of the alcohol–chloride interaction from the number of such interactions formed in solution, the binding energy of the isolated ROH···$Cl^-$ complex was computed for each alcohol (Supplementary Table 4).

**Supplementary Table 4.** Binding energy of the ROH···$Cl^-$ hydrogen-bonded complex. The chloride reference cancels exactly in the alcohol-to-alcohol differences given in the final column.

| **Alcohol** | **Binding energy in chlorobenzene (eV)** | **Binding energy in DMF (eV)** | **Relative to ethanol (eV)** |
|---|---|---|---|
| Methanol | −0.356 | −0.267 | +0.002 |
| Ethanol | −0.358 | −0.269 | 0.000 |
| 1-Propanol | −0.386 | −0.294 | −0.028 |
| 1-Butanol | −0.383 | −0.290 | −0.025 |

The intrinsic hydrogen-bond strength is almost invariant across the series. Methanol and ethanol differ by 2 meV, 1-propanol and 1-butanol by 3 meV, and the two pairs are separated by only ≈27 meV — approximately one $k^bT$ in total across four carbon numbers. This is expected, since the interaction is dominated by the hydroxyl group, which is common to all four molecules.

The contrast with the molecular dynamics is the informative result. The number of alcohol–chloride hydrogen bonds formed in solution differs by more than a factor of three between ethanol and 1-propanol, while the strength of an individual such bond differs by 28 meV. The difference in alcohol–chloride association therefore does not arise from the intrinsic strength of the hydrogen bond, but more likely from how alcohol molecules are accommodated in the solvation environment around the ionic precursor; the present simulations do not resolve which factors dominate.

### 2.3.3 Hydrogen bonding to the PEA ammonium group

The growth mechanism invokes alcohol hydrogen bonding to chloride and to the ammonium group of $PEA^+$. These are distinct interactions in which the alcohol plays opposite roles: it is the donor to chloride (O–H···$Cl^-$) and the acceptor from ammonium (N–H···O). The ammonium interaction was therefore evaluated separately, using the complete phenethylammonium cation so that the steric environment created by the phenethyl group is represented.

**Supplementary Table 5.** Binding energy and geometry of the $PEA^{+}\cdots HO–R$ hydrogen-bonded complex, in which the alcohol acts as the hydrogen-bond acceptor.

| **Alcohol** | **Binding energy in chlorobenzene (eV)** | **Binding energy in DMF (eV)** | **N···O (Å)** | **H···O (Å)** |
|---|---|---|---|---|
| Methanol | −0.378 | −0.339 | 2.77 | 1.73 |
| Ethanol | −0.396 | −0.291 | 2.78 | 1.74 |
| 1-Propanol | −0.467 | −0.299 | 2.78 | 1.74 |
| 1-Butanol | −0.396 | −0.295 | 2.78 | 1.73 |

The hydrogen-bond geometry is essentially identical across the series (N···O 2.77–2.78 Å, H···O 1.73–1.74 Å in chlorobenzene), indicating a common binding motif. Methanol, ethanol and 1-butanol bind within 19 meV of one another. The 1-propanol value lies 71–89 meV below the others although its hydrogen-bond geometry is indistinguishable from theirs. This is a reference-state artefact: in the 1-propanol complex the phenyl ring of $PEA^{+}$ is rotated by about 90° relative to the isolated cation used as the reference ($PEA^{+}$ heavy-atom RMSD 0.50 Å, against ≤0.05 Å for the other complexes), so conformational relaxation of the cation is counted as binding. The same rotation occurs for methanol in the DMF continuum (RMSD 0.46 Å) and accounts for the methanol value in that column. Neither value should be interpreted as a chain-length effect.

Taken with Supplementary Table 4, the conclusion is consistent for both interactions named in the growth mechanism: the strength of the alcohol hydrogen bond is effectively independent of chain length, varying by no more than about 30 meV across C1 to C4 once the conformational artefacts above are excluded. Since the number of such contacts observed in solution varies by more than a factor of three, differences in alcohol association between the alcohols are not explained by the strength of the individual interaction and more likely reflect the solvation environment around the precursor.

One qualitative observation from the same trajectories is robust across all four systems: where an alcohol occupies the ammonium site, the number of $N–H\cdots Cl^{-}$ contacts falls to 2.9–3.8, whereas where it does not the count is 4.9–5.0. The alcohol and chloride compete for the ammonium group, and the alcohol displaces chloride when present.

## 2.4 Methanol is a distinct crystallisation regime

Methanol is excluded from the chain-length analysis above, for both an experimental and a computational reason.

Experimentally, methanol does not produce discrete ultrathin crystals. The optical micrographs show a continuous mat of aggregated, unfaceted and polydisperse particulate, whereas ethanol under otherwise identical conditions yields separated, sharply faceted platelets. A lateral dimension extracted from the methanol sample does not measure the same physical quantity as one extracted from the other alcohols, and the two should not be placed on a common axis.

Computationally, methanol is anomalous in a specific way. It associates with the ionic precursor approximately three times more strongly than any other alcohol (enrichment 18.2 against 3.8–5.7; 3.70 hydrogen bonds against 0.57–1.98), yet its ability to displace DMF from copper (+0.040 eV) is indistinguishable from that of ethanol (+0.037 eV), and its intrinsic hydrogen-bond strengths to chloride and to the ammonium group are likewise within 2 and 19 meV of ethanol's. Methanol is therefore not simply the extreme member of a monotonic series: it crowds the precursor far more heavily while competing no more effectively for the metal coordination site or for the chloride.

Taken together with the observed morphology, this is consistent with wholesale destabilisation of the precursor solution, producing rapid and uncontrolled precipitation rather than the controlled nucleation and lateral growth obtained with the longer alcohols.

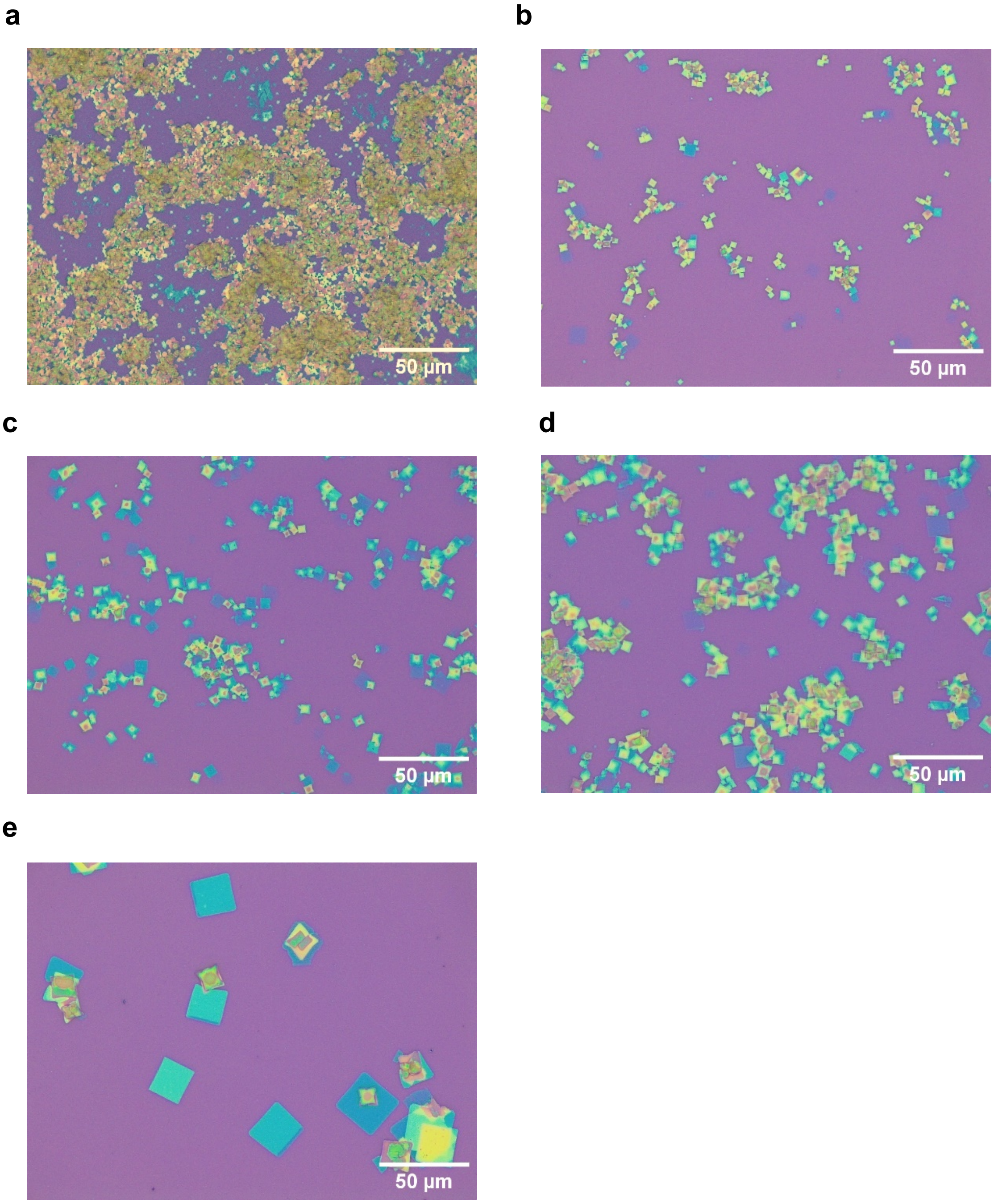

**Supplementary Figure 1.** Optical microscopy images of $(PEA)_2CuCl_4$ layers synthesized with different alcohols in the ternary solvent system (CB: Alcohol: DMF = 3200: 400: 10). The alcohols used are (a) methanol (b) ethanol, (c) 2-propanol, (d) 1-propanol, and (e) 1-butanol.

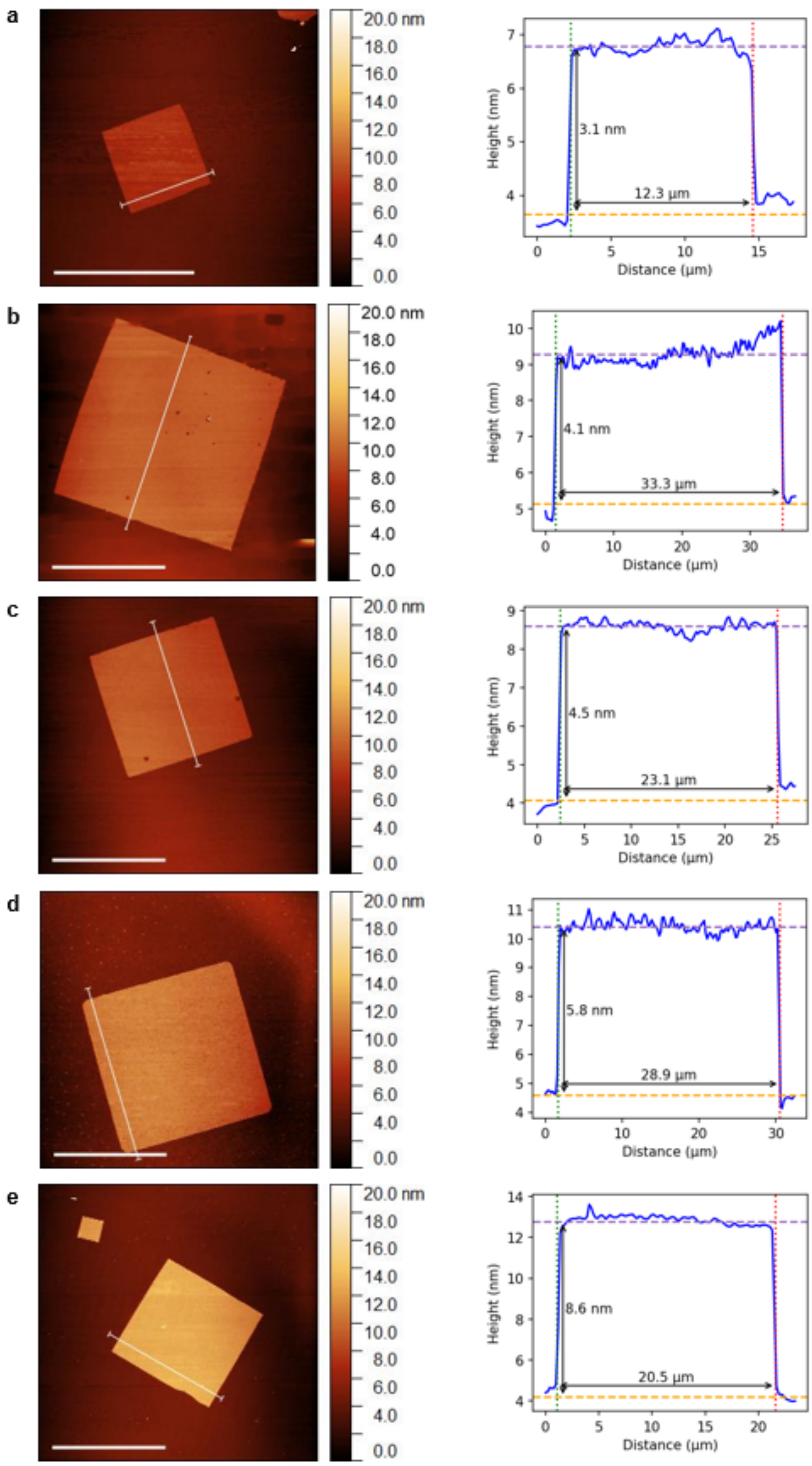

**Supplementary Figure 2.** AFM topography scans of ultra-thin $(PEA)_2CuCl_4$ layers. Images show isolated flakes with (a) monolayer (~3.1 nm), (b-c) 2-layer (~4.1, 4.5 nm), (d) 3-layer (~5.8 nm), and (e) 4-layer (~8.6 nm) thickness. (Scale bar: 20 μm)

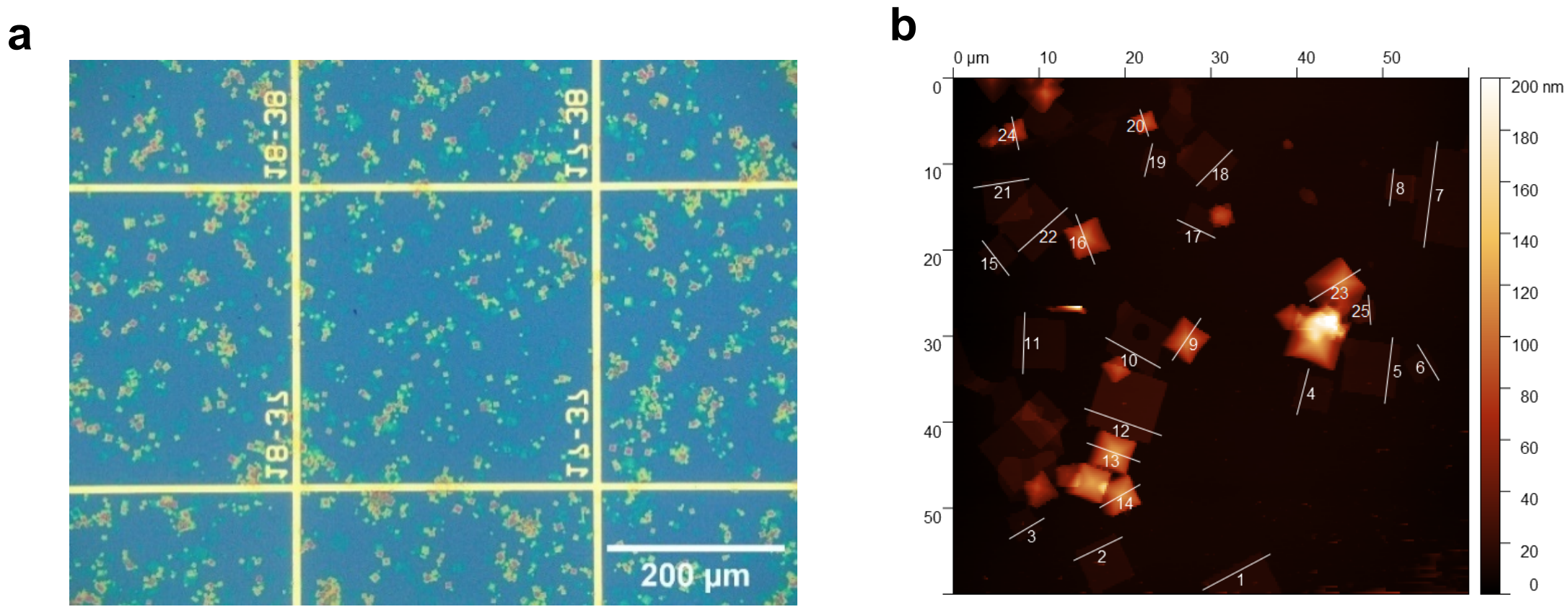

**Supplementary Figure 3**. (a) Optical microscopy image of $(PEA)_2CuCl_4$ flakes synthesized using the ternary solvent system (CB:alcohol:DMF = 3200:400:10 µL) on a pre-patterned $Si/SiO_2$ substrate containing a fiducial marker grid. The grid enables precise localization of individual flakes for subsequent AFM analysis. (b) Representative AFM topography image showing selected flakes and line profiles used for dimensional measurements.

### Dimensional analysis of $(PEA)_2CuCl_4$ flakes

Ultrathin $(PEA)_2CuCl_4$ flakes were prepared on pre-patterned $Si/SiO_2$ substrates containing $250 \times 250$ $\mu m^2$ fiducial grid squares. Following synthesis, optical microscopy was used to assess the overall flake distribution and identify representative regions for AFM analysis. Multiple flakes from different grid regions were subsequently examined to account for spatial variations in flake size and thickness.

AFM height images were analysed using Gwyddion software. For each flake, cross-sectional line profiles were drawn across the flake edges, as illustrated in Supplementary Figure 3. The flake thickness was determined from the height difference between the $SiO_2$ substrate and the flake terrace, while the lateral width was obtained from the same profile. Measurements from all analysed flakes were compiled to generate the thickness-versus-width distributions presented in Figure 2i-m of the main text.

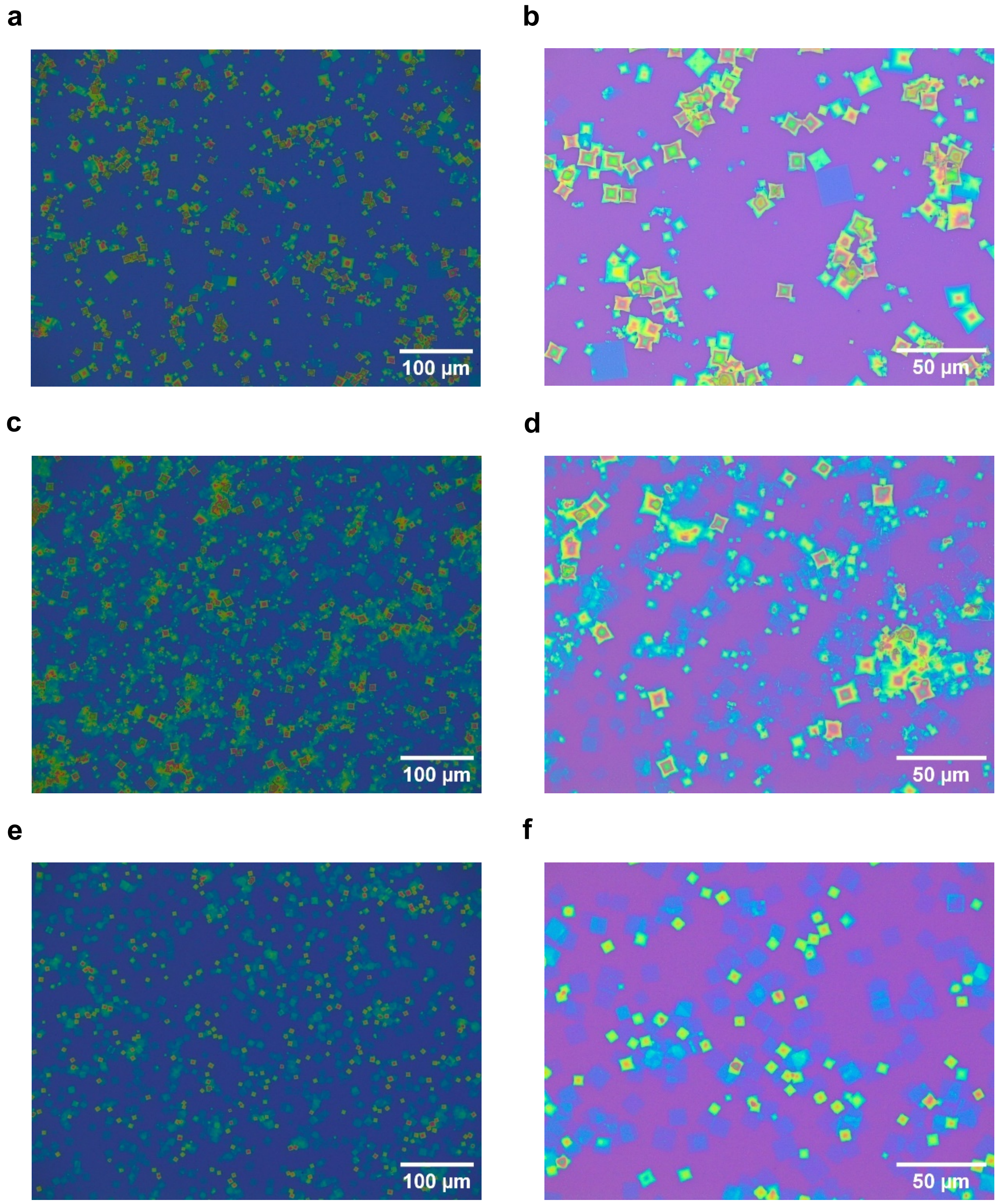

**Supplementary Figure 4.** Optical microscopy images of $(PEA)_2CuCl_4$ layers synthesized with varying concentrations of octadecylamine (ODA) additives: (a, b) Without ODA additives, (c, d) $2.8 \times 10^{-7}$ M ODA, and (e, f) $5.6 \times 10^{-7}$ M ODA.

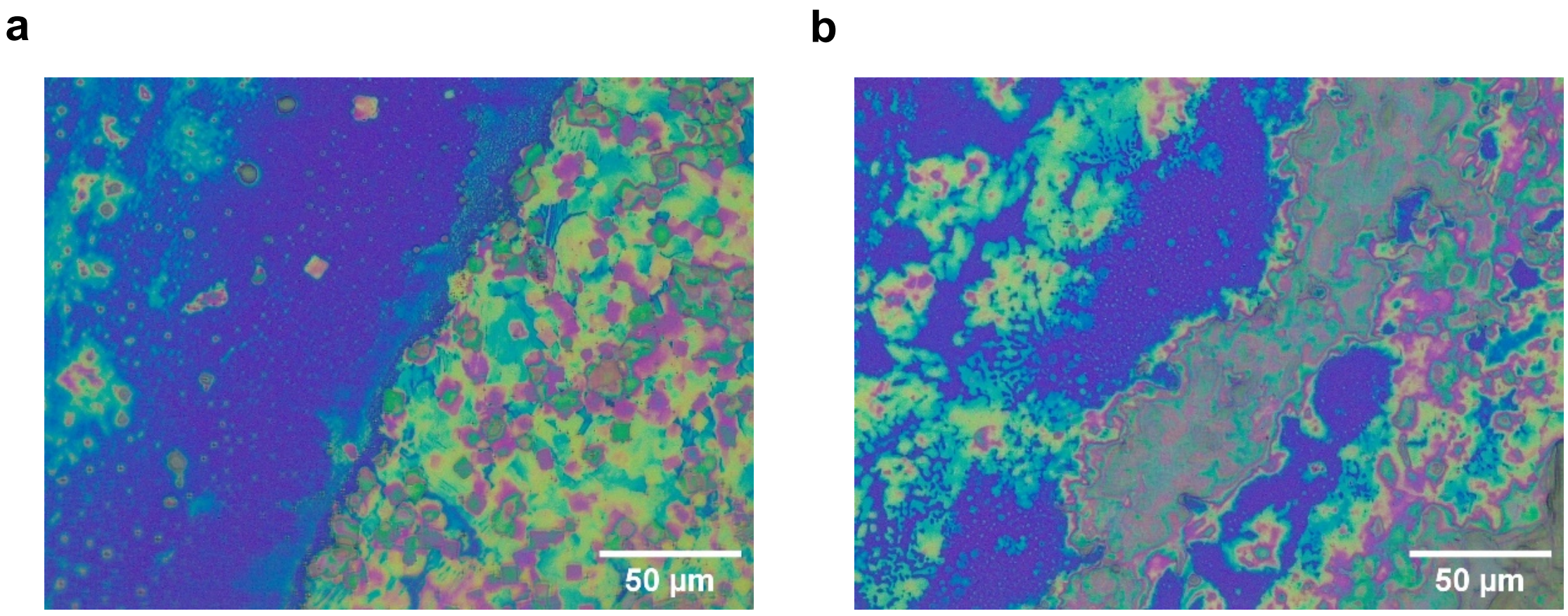

**Supplementary Figure 5.** Optical microscopy images of $(PEA)_2CuCl_4$ layers synthesized using long-chain alcohols. The layers were prepared using the standard ternary solvent system (CB: Alcohol: DMF 3200:400:10) with (a) pentanol and (b) hexanol.

Supplementary Text

Based on our experimental observations, increasing the alcohol chain length beyond C4 did not further improve monolayer formation (Supplementary Figure 5). This limitation arises from the differential evaporation rates within the ternary solvent mixture. Short-chain alcohols such as methanol, ethanol, and propanol evaporate more rapidly than chlorobenzene and dimethylformamide (DMF) due to their higher vapor pressures.[1] As these alcohols evaporate prematurely, the solvent mixture becomes enriched in chlorobenzene and DMF. During the final annealing stages, this residue of slower-evaporating solvents promotes re-agglomeration of ultrathin layers through enhanced van der Waals forces and π–π interactions between the phenethylammonium spacers.[2] In contrast, butanol, with a vapour pressure (~0.9 kPa) lower than chlorobenzene (~1.6 kPa at 25 °C), evaporates more slowly and therefore remains in the ternary solvent mixture throughout annealing. This behavior helps preserve the target 8:1 chlorobenzene-to-alcohol ratio during processing, thereby supporting the stable growth of large, ultrathin layers.

However, when the alcohol chain length increases further, solvents such as pentanol (boiling point 137.5 °C) and hexanol (157 °C) introduce new challenges. Their substantially lower vapor pressures relative to chlorobenzene cause them to persist in the mixture even after chlorobenzene has evaporated. Additionally, their higher viscosities hinder uniform spreading of the precursor solution during drop-casting and promote aggregation, ultimately suppressing 2D layer formation.

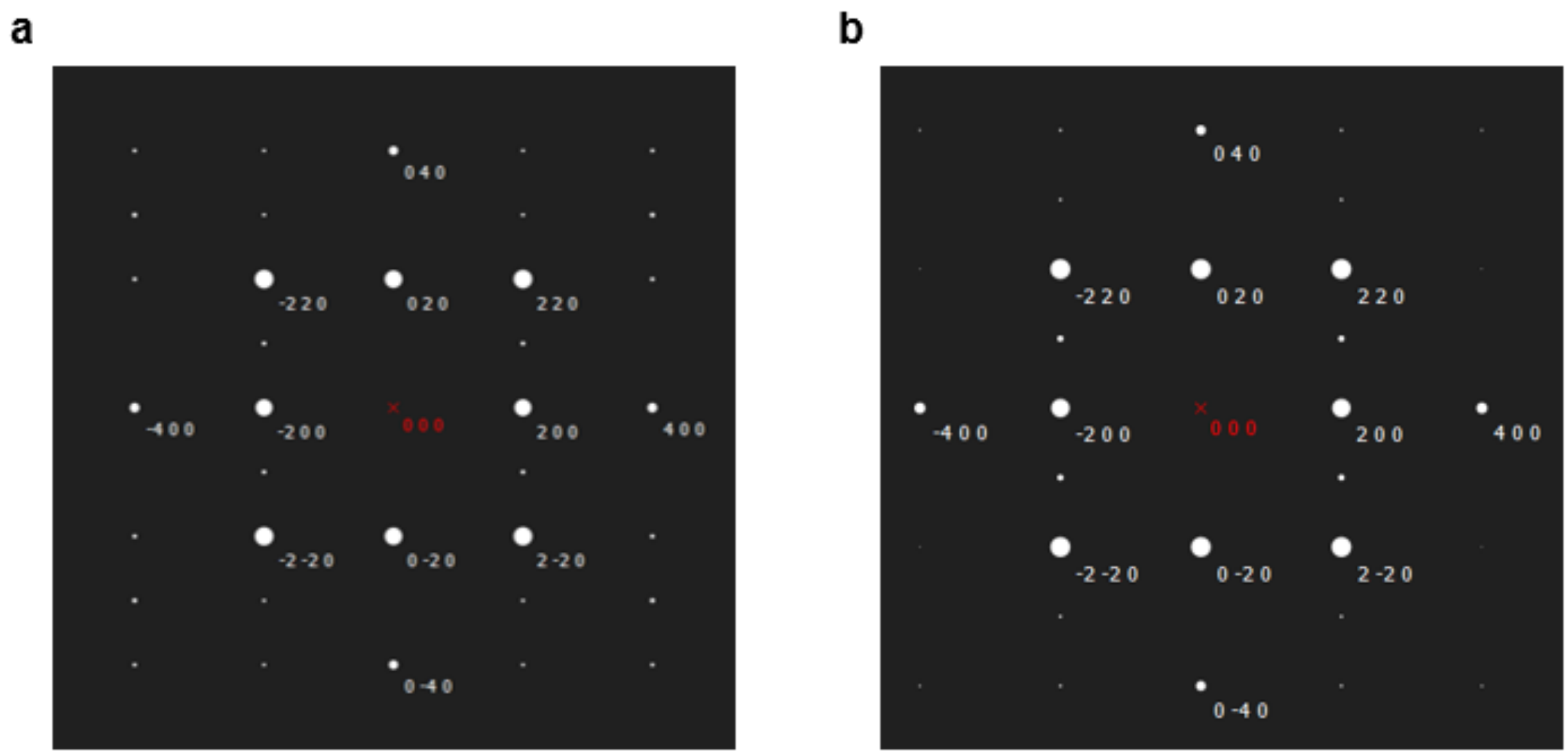

**Supplementary Figure 6.** Simulated selected-area electron diffraction (SAED) patterns for (a) $(PEA)_2CuCl_4$ and (b) $(PEA)_2MnCl_4$. Simulations were performed using the ReciPro crystallographic software [3] with the following parameters: a parallel electron beam with 200 kV accelerating voltage and intensity calculations based on kinematical scattering with excitation errors.

**Supplementary Table 6.** Comparison of measured SAED and simulated d-spacings for $(PEA)_2CuCl_4$. The simulated and experimental values show excellent agreement with negligible deviation.

| Simulation | | | | TEM-SAED measurement | | |
|---|---|---|---|---|---|---|
| h | k | l | d (Å) | 2*(1/d) | 1/d | d (Å) |
| 0 | 2 | 0 | 3.63 | 5.46 | 2.73 | 3.66 |
| 2 | 0 | 0 | 3.60 | 5.52 | 2.76 | 3.62 |
| 2 | 2 | 0 | 2.56 | 7.71 | 3.855 | 2.59 |
| 0 | 4 | 0 | 1.82 | 10.86 | 5.43 | 1.84 |
| 4 | 0 | 0 | 1.80 | 11 | 5.5 | 1.82 |

**Supplementary Table 7.** Comparison of measured SAED and simulated d-spacings for $(PEA)_2MnCl_4$. The simulated and experimental values show excellent agreement with negligible deviation.

| **Simulation** | | | | **TEM-SAED measurement** | | |
|---|---|---|---|---|---|---|
| **h** | **k** | **l** | **d (Å)** | **2*(1/d)** | **1/d** | **d (Å)** |
| 0 | 2 | 0 | 3.65 | 5.52 | 2.76 | 3.62 |
| 2 | 0 | 0 | 3.60 | 5.62 | 2.81 | 3.56 |
| 2 | 2 | 0 | 2.56 | 7.88 | 3.94 | 2.54 |
| 0 | 4 | 0 | 1.82 | 11.08 | 5.54 | 1.81 |
| 4 | 0 | 0 | 1.80 | 11.34 | 5.67 | 1.76 |

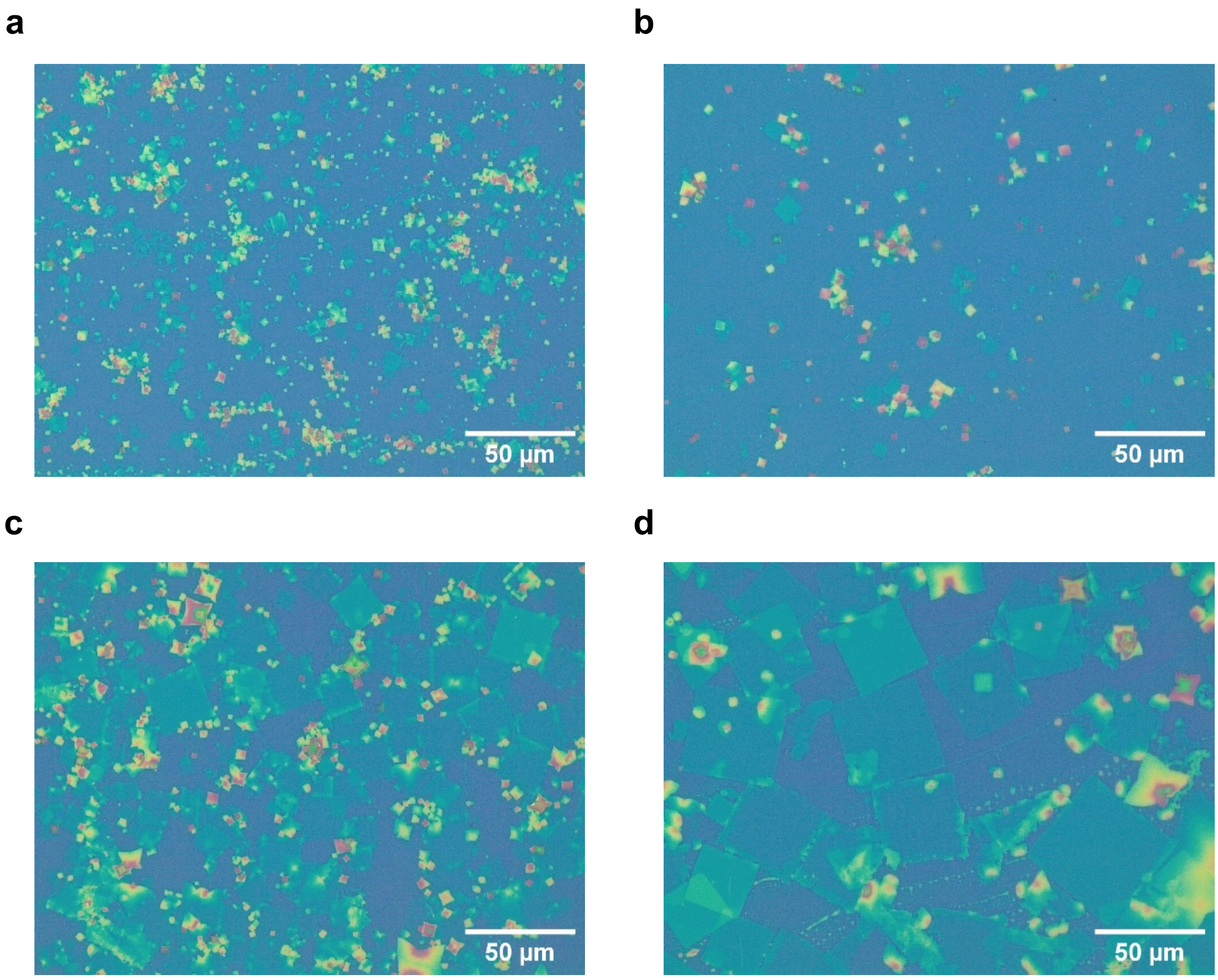

**Supplementary Figure 7.** Optical microscopy images of $(FPEA)_2CuCl_4$ layers synthesized with different alcohols in the ternary solvent system (CB: Alcohol: DMF = 3100: 500: 10). The alcohols used are (a) ethanol, (b) 2-propanol, (c) 1-propanol, and (d) 1-butanol.

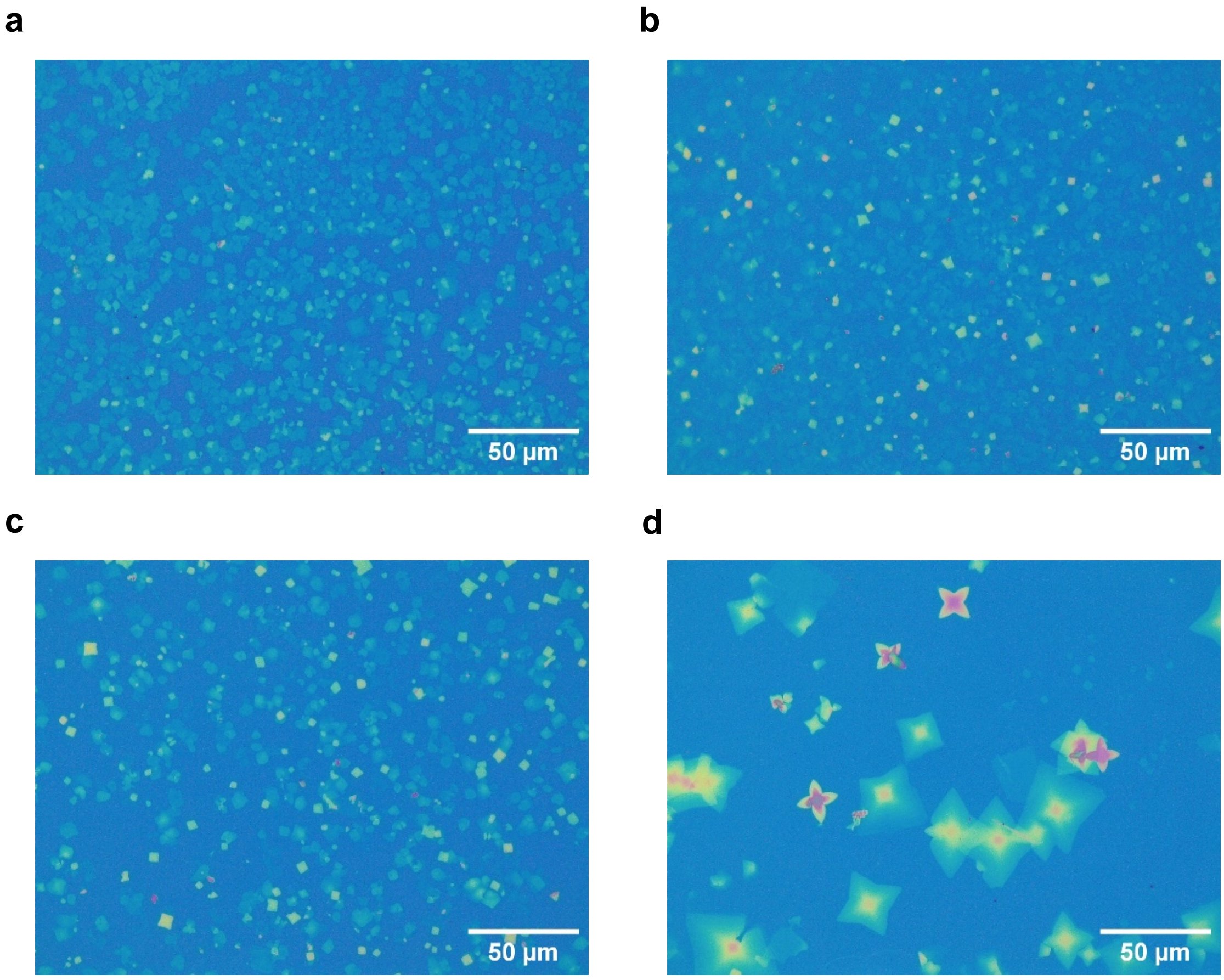

**Supplementary Figure 8.** Optical microscopy images of $(PentA)_2CuCl_4$ layers synthesized with different alcohols in the ternary solvent system (CB: Alcohol: DMF = 3300: 308: 2). The alcohols used are (a) ethanol, (b) 2-propanol, (c) 1-propanol, and (d) 1-butanol.

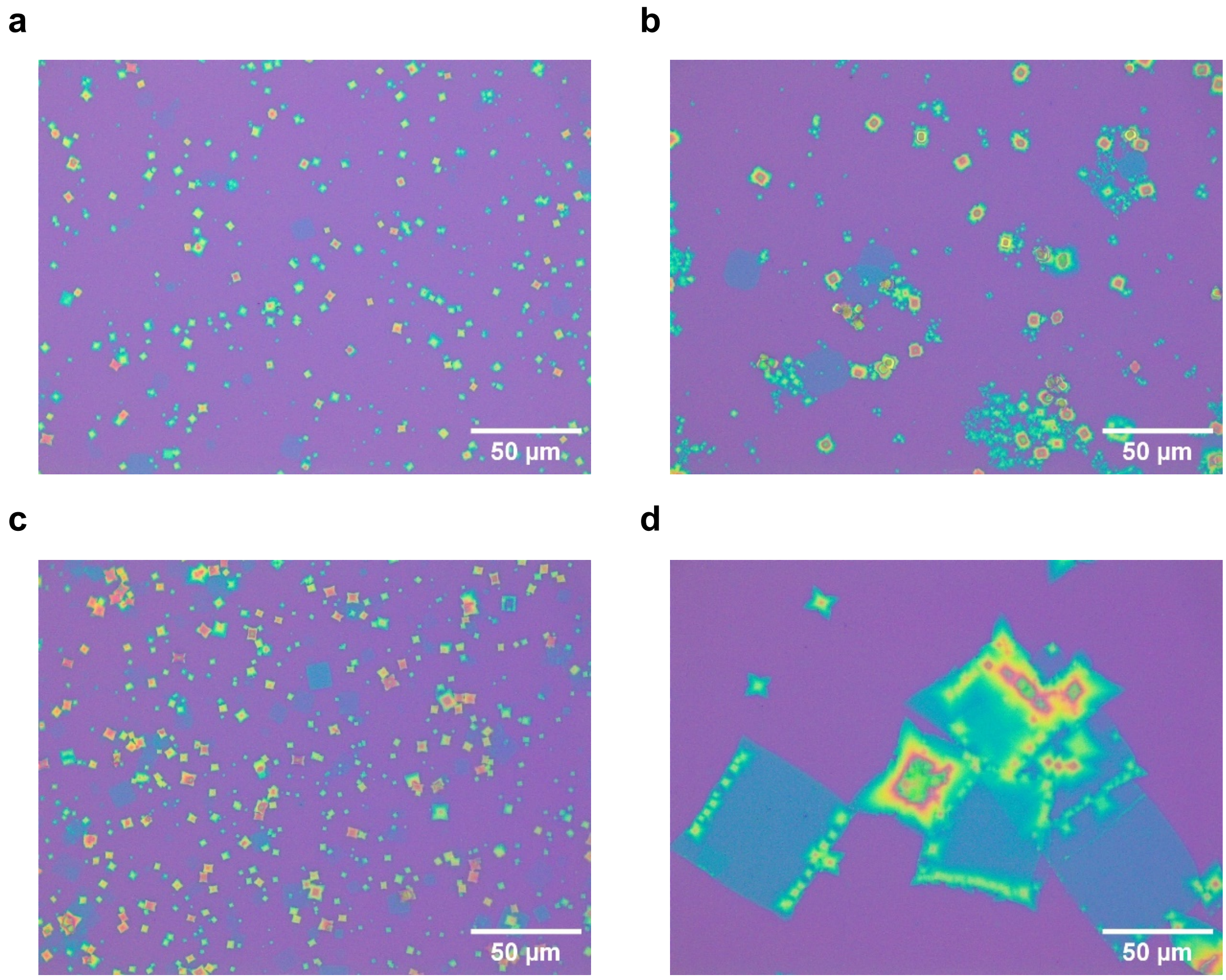

**Supplementary Figure 9.** Optical microscopy images of $(PEA)_2MnCl_4$ layers synthesized with different alcohols in the ternary solvent system (CB: Alcohol: DMF = 3100: 500: 10). The alcohols used are (a) ethanol, (b) 2-propanol, (c) 1-propanol, and (d) 1-butanol.

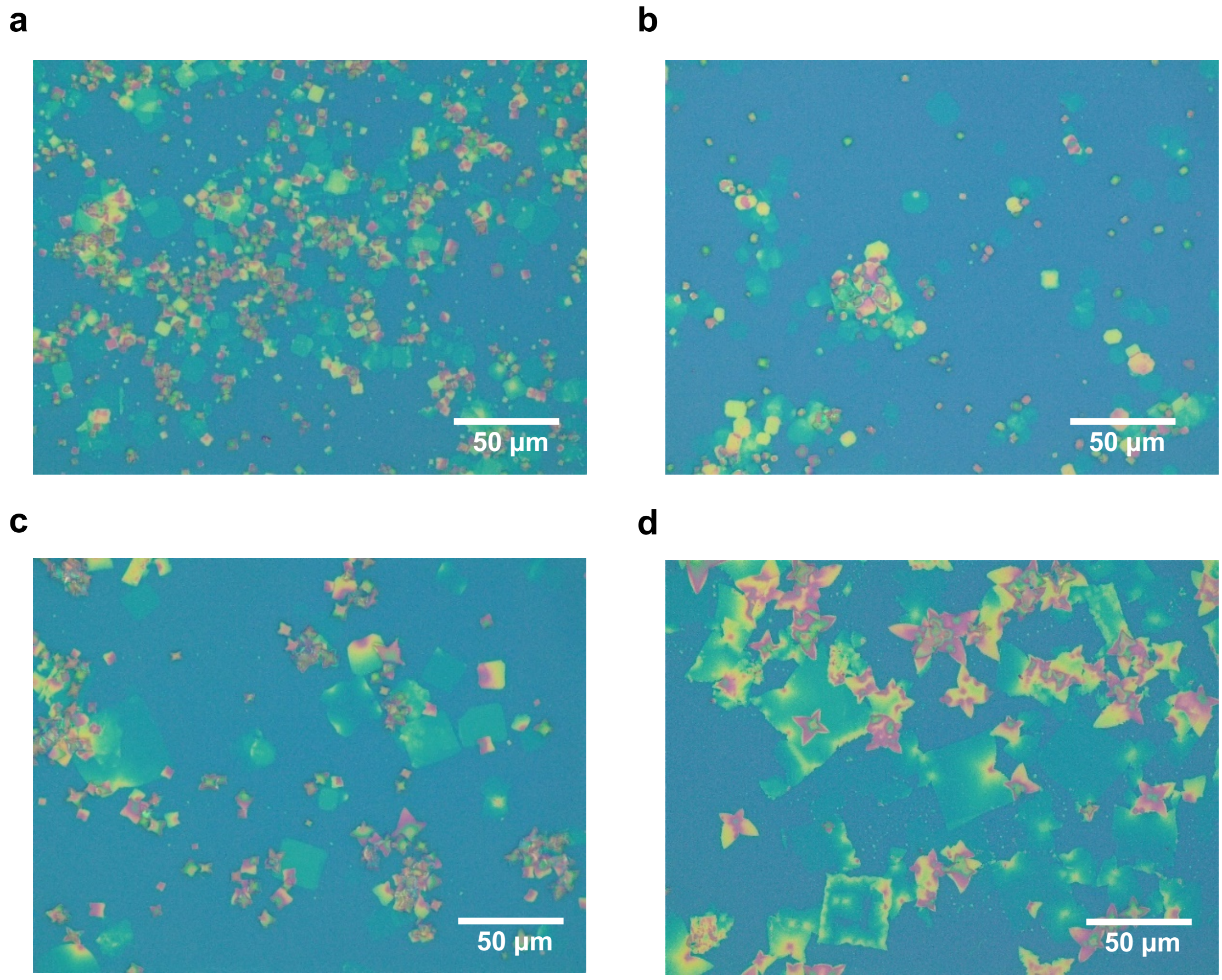

**Supplementary Figure 10.** Optical microscopy images of $(FPEA)_2MnCl_4$ layers synthesized with different alcohols in the ternary solvent system (CB: Alcohol: DMF = 3000: 600: 10). The alcohols used are (a) ethanol, (b) 2-propanol, (c) 1-propanol, and (d) 1-butanol.

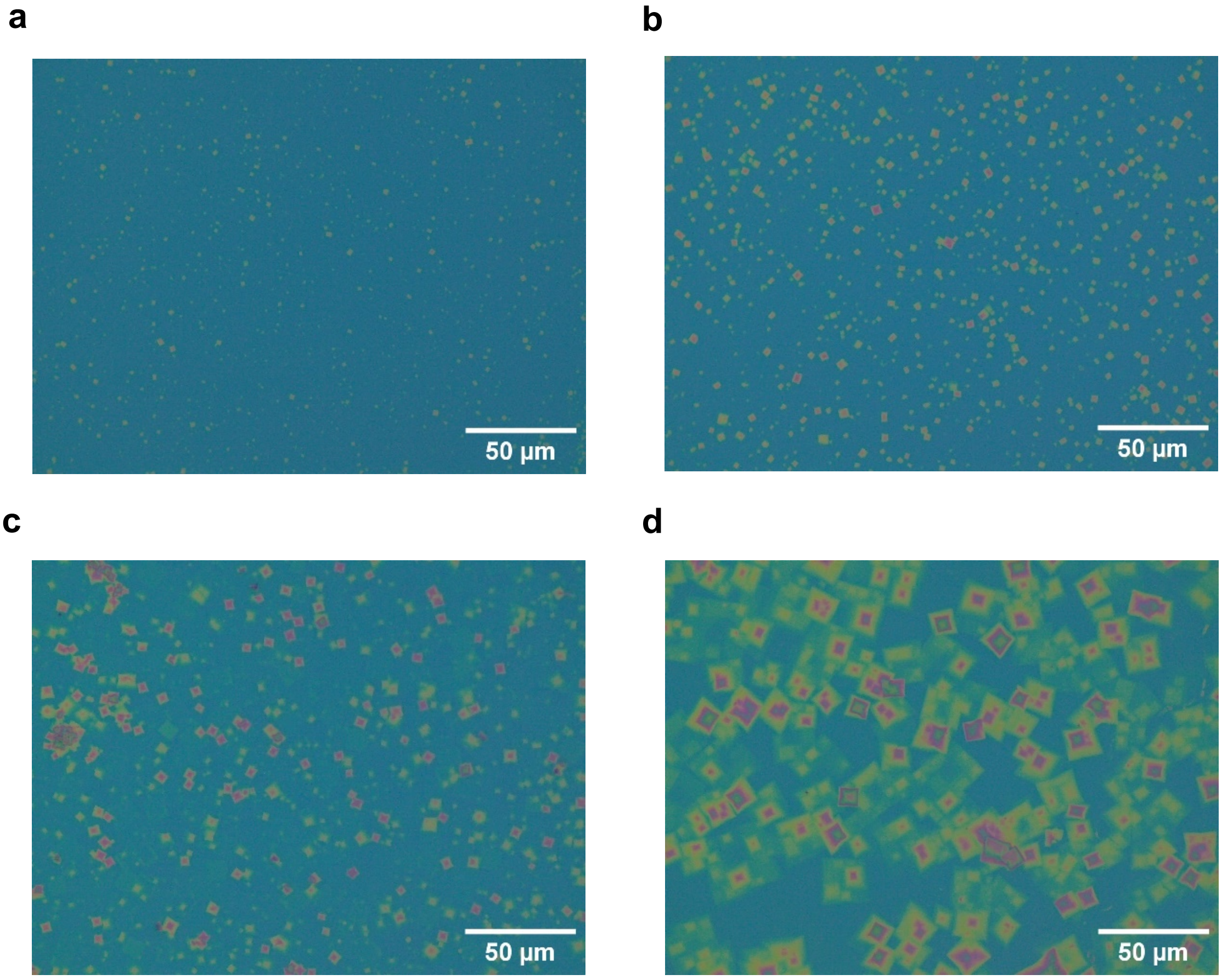

**Supplementary Figure 11.** Optical microscopy images of $(PentA)_2MnCl_4$ layers synthesized with different alcohols in the ternary solvent system (CB: Alcohol: DMF = 3300: 300: 10). The alcohols used are (a) ethanol, (b) 2-propanol, (c) 1-propanol, and (d) 1-butanol.

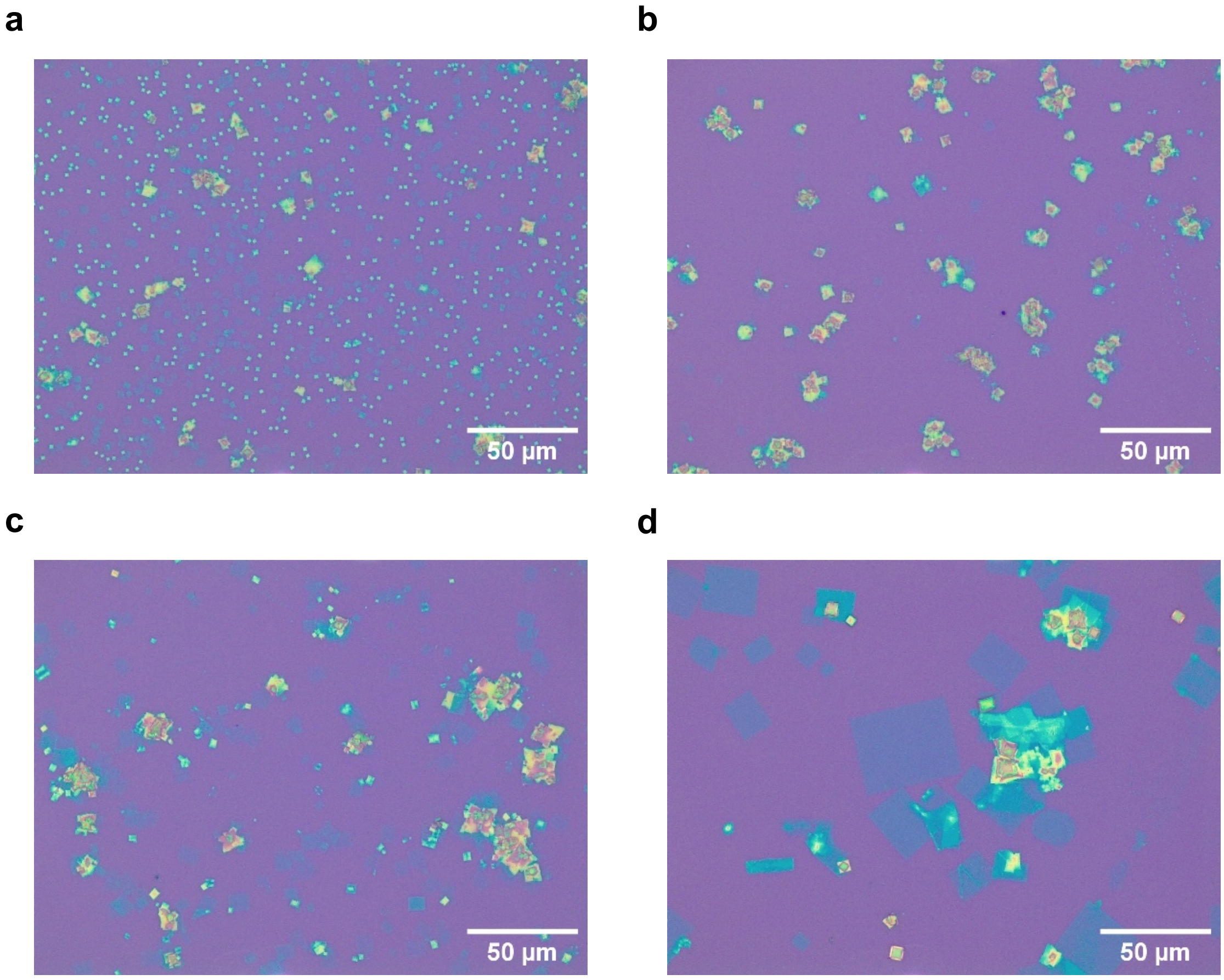

**Supplementary Figure 12.** Optical microscopy images of $(PEA)_2PbBr_4$ layers synthesized with different alcohols in the ternary solvent system (CB: Alcohol: DMF = 3000: 600: 30). The alcohols used are (a) ethanol (b) 2-propanol, (c) 1-propanol, and (d) 1-butanol.

**References**

1. Li, X. *et al.* Azeotropic Binary Solvent Mixtures for Preparation of Organic Single Crystals. *Advanced Functional Materials* **19**, 3610–3617 (2009).
2. Gao, Y., Dong, X. & Liu, Y. Recent Progress of Layered Perovskite Solar Cells Incorporating Aromatic Spacers. *Nano-Micro Letters* **15**, 1–19 (2023).
3. Seto, Y. & Ohtsuka, M. ReciPro: free and open-source multipurpose crystallographic software integrating a crystal model database and viewer, diffraction and microscopy simulators, and diffraction data analysis tools. *J Appl Cryst* **55**, 397–410 (2022).